\documentclass{pasa}%

\usepackage{graphicx}
\usepackage{amsmath}	
\usepackage{amssymb}	
\usepackage{color}

\DeclareSymbolFont{UPM}{U}{eur}{m}{n}
\SetSymbolFont{UPM}{bold}{U}{eur}{b}{n}
\DeclareSymbolFont{AMSa}{U}{msa}{m}{n}
\DeclareMathSymbol{\umu}{0}{UPM}{"16}

\newcommand\arcmin{\hbox{$^\prime$}}
\newcommand\arcsec{\hbox{$^{\prime\prime}$}}

\title[GravityCam]{GravityCam: Wide-Field High-Resolution High-Cadence Imaging Surveys in the Visible from the Ground}

\author[C. Mackay et al.]{C. Mackay$^{1}$\thanks{E-mail: cdm<at>ast.cam.ac.uk},
M.~Dominik$^{2}$,
I.A.~Steele$^{3}$,
C.~Snodgrass$^{4}$,
U.G.~J{\o}rgensen$^{5}$,
J.~Skottfelt$^{4}$,
K.~Stefanov$^{4}$,
B.~Carry$^{6}$,
F.~Braga-Ribas$^{7}$,
A.~Doressoundiram$^{8}$,
V.D.~Ivanov$^{9,10}$,
P.~Gandhi$^{11}$,
D.F.~Evans$^{12}$,
M.~Hundertmark$^{13}$,
S.~Serjeant$^{4}$,
S.~Ortolani$^{14,15}$
\affil{$^{1}$Institute of Astronomy, University of Cambridge, Cambridge, CB3 0HA, United Kingdom}
\affil{$^{2}$Centre for Exoplanet Science, SUPA School of Physics \& Astronomy, University of St Andrews, North Haugh, St Andrews, KY16 9SS, United Kingdom}
\affil{$^{3}$Astrophysics Research Institute, Liverpool John Moores University, Liverpool CH41 1LD, United Kingdom}
\affil{$^{4}$School of Physical Sciences, The Open University, Milton Keynes, MK7 6AA, United Kingdom}
\affil{$^{5}$Niels Bohr Institute \& Centre for Star and Planet Formation, University of Copenhagen, {\O}ster Voldgade 5, 1350 Copenhagen, Denmark}
\affil{$^{6}$Universit\'{e} C\^{o}te d'Azur, Observatoire de la C\^{o}te d'Azur, CNRS, Laboratoire Lagrange, France}
\affil{$^{7}$Federal University of Technology - Paran\'a (UTFPR / DAFIS), Curitiba, Brazil}
\affil{$^{8}$Observatoire de Paris-LESIA, 5 Place Jules Janssen, Meudon Cedex 92195, France}
\affil{$^{9}$European Southern Observatory, Ave. Alonso de C\'ordova 3107,Vitacura, Santiago, Chile}
\affil{$^{10}$European Southern Observatory, Karl-Schwarzschild-Str. 2, 85748 Garching bei M\"unchen, Germany}
\affil{$^{11}$Department of Physics \& Astronomy, University of Southampton, Highfield, Southampton, SO17 1BJ, United Kingdom}
\affil{$^{12}$Astrophysics Group, Keele University, Staffordshire, ST5 5BG, United Kingdom}
\affil{$^{13}$Astronomisches Rechen-Institut, Zentrum f\"ur Astronomie der Universit\"at Heidelberg (ZAH), 69120 Heidelberg, Germany}
\affil{$^{14}$Dipartimento di Fisica e Astronomia, Universit\`a degli Studi di Padova,
 Vicolo dell'Osservatorio 3, 35122, Padova, Italy}
\affil{$^{15}$Osservatorio Astronomico di Padova, INAF, Vicolo dell'Osservatorio 5, 35122, Padova, Italy}
}

\jid{PASA}
\doi{10.1017/pas.\the\year.xxx}
\jyear{\the\year}

\usepackage{aas_macros}
\usepackage{hyperref} 
\hypersetup{colorlinks,citecolor=blue,linkcolor=blue,urlcolor=blue}

\begin{document}

\begin{frontmatter}
\maketitle

\begin{abstract}
\mbox{GravityCam} is a new concept of ground-based imaging instrument capable of delivering significantly sharper images from the ground than is normally possible without adaptive optics.  Advances in optical and near infrared imaging technologies allow images to be acquired at high speed without significant noise penalty.  Aligning these images before they are combined can yield a 2.5--3 fold improvement in image resolution.  By using arrays of such detectors, survey fields may be as wide as the telescope optics allows. Consequently, GravityCam enables both wide-field high-resolution imaging and high-speed photometry.
We describe the instrument and detail its application to provide demographics of planets and satellites down to Lunar mass (or even below) across the Milky Way. \mbox{GravityCam} is also suited to improve  the quality of weak shear studies of dark matter distribution in distant clusters of galaxies and multiwavelength follow-ups of background sources that are strongly lensed by galaxy clusters. The photometric data arising from an extensive microlensing survey will also be useful for asteroseismology studies, while \mbox{GravityCam} can be used to monitor fast multiwavelength flaring in accreting compact objects, and promises to generate a unique data set on the population of the Kuiper belt and possibly the Oort cloud.

\end{abstract}

\end{frontmatter}



\vspace*{3cm}

\section{Introduction}

Astronomers have learned a great deal about the Universe using data from large-scale direct imaging surveys of the sky: from the original Photographic Sky Surveys carried out by dedicated Schmidt telescopes in both North and South Hemisphere \citep{Reid1993}  to the Sloan Digital Sky Survey made with a 2.5 m telescope with a very wide field to provide high quality digital data of a significant part of the northern sky \citep{Gunn+2006}. With the 4.1m VISTA (VLT Infrared Survey Telescope for Astronomy; \citealt{Emerson:VISTA}) and the 2.5m VST (VLT Survey Telescope; \citealt{VST}), ESO began operating two new wide field survey facilities  that are imaging the complete Southern sky, introducing a new mode of ESO Public surveys that make reduced data and a number of high-level data products available to the community. More recently, the Dark Energy Survey (DES -- https://www.darkenergysurvey.org/) uses a modified 4 m class telescope in Chile with the large detector area to take deep images of a substantial part of the southern hemisphere.  More and bigger surveys are planned in the future, such as the Large Synoptic Survey Telescope (LSST; \citealt{LSST}), which is in an early stage of construction and uses an 8 m class telescope also located in Chile with a 3.5$^\circ$ field of view to image large areas of the sky repeatedly with relatively short exposures.  Surveying much of the sky every few nights can lead to the detection of exploding supernovae in distant galaxies as well as earth-approaching objects.

It is the increasing sophistication of digital, principally charge coupled device (CCD), technology, that has enabled these surveys to progress by generating large volumes of data so easily.  CCDs provided high-quality repeatable electronic detectors with detective quantum efficiency approaching 100\% at best, broad spectral response, the capability of integrating signals accurately over long periods of time as well as being available in large formats.  What has not changed in the last 70 years is the capacity of these surveys to deliver images any sharper than those of the original photographic sky survey.  Although the highest redshift probed by wide-field surveys have increased from $z \sim 0.2$ to $z >11$, the vast majority of the most distant objects are essentially unresolved by ground-based telescopes.  Even within our own Galaxy there are many regions where stars are so close together on the sky as to be badly confused.  Advances in adaptive optics technologies have allowed higher resolution images to be obtained over very small fields of view, a few arcseconds at best and therefore of no help in delivering sharper images in wide-field surveys.  It is this deficiency that \mbox{GravityCam} is intended to overcome. \mbox{GravityCam} is not intended to produce diffraction limited images, but simply ones that are sharper than can normally be obtained from the ground. This will particularly allow imaging large areas of the inner Milky Way and nearby galaxies such as the Magellanic clouds with unprecedented angular resolution. These areas are abundant of bright stars necessary for image alignment. Moreover, by operating \mbox{GravityCam} at frame rates $> 10~\mbox{Hz}$, surveys at very high cadence will be enabled.

There are several key scientific programmes that will benefit substantially from such an instrument and are described below in more detail.  They include the detection of extra-solar planets and satellites down to Lunar mass by surveying tens of millions of stars in the bulge of our own Galaxy, and the detection and mapping of the distribution of dark matter in distant clusters of galaxies by looking at the distortions in galaxy images.  \mbox{GravityCam} can provide a unique new input by surveying millions of stars with high time resolution to enable asteroseismologists to better understand the structure of the interior of those stars. It can also allow a detailed survey of Kuiper belt objects via stellar occultations.

Efforts to procure funding for GravityCam are underway by the GravityCam team
following the first international GravityCam workshop in June 2017 at the Open University.

This article provides a detailed account of the technology and potential envisaged science cases of GravityCam, following and substantially elaborating on the brief overview previously provided by \citet{SPIE:GravityCam}.
We give a technical overview of the instrument in Section~\ref{Sect:outline}, whereas Sections~\ref{Sect:Microlensing} to~\ref{Sect:other} present some examples of scientific breakthroughs that will be possible with GravityCam. Sections~\ref{Sect:Instrument} to~\ref{Sect:sensitive} give an in-depth description of the instrument, before we provide a summary and conclusions in Section~\ref{Sect:Conclusions}.

\section{Technical outline of GravityCam}
\label{Sect:outline}

\subsection{High angular resolution with lucky imaging}

The image quality of conventional integrating cameras is usually constrained by atmospheric turbulence
characteristics.  Such turbulence has a power spectrum that is strongest on the largest scales \citep{Fried}. One of the most straightforward ways to improve the angular resolution of images on a telescope therefore is to take images rapidly (in the 10--30 Hz range) and use the position of a bright object in the field to allow its offset relative to some mean to be established.  
This technique almost completely eliminates the tip tilt distortions caused by atmospheric turbulence. 
The next level of disturbance of the wavefront entering the telescope is defocus.  Using the same procedure described above but only adding the best and sharpest images together is the method known as Lucky Imaging \citep{Mackay+2004}. Even more demanding selection can ultimately give even higher factors (e.g. \citealt{Baldwin+2008}), but the resolution resulting from less strict selection will often meet the requirements of many scientific applications.

Lucky Imaging is already well established as an astronomical technique with over 350 papers already published mentioning ``Lucky Imaging'' in the abstract, including over 40 from the Cambridge group, which include specific examples of results obtained with the systems \citep{Law+2006,Scardia+2007,Mackay+2008,Law+2009,Faedi+2013,Mackay2013}.

Lucky Imaging works very well for small diameter telescopes, yielding resolution similar to Hubble ($\sim\,0.1\arcsec$) on Hubble size ($\sim\,2.5~\mbox{m}$) telescopes.  With larger telescopes the chance of obtaining high-resolution images become smaller.  Although it would be very convenient to achieve even higher resolution on bigger telescopes without any more effort, in practice other techniques such as combining Lucky Imaging with low order adaptive optics need to be used which are beyond the scope of this paper \citep{Law+2009}.  However we propose siting \mbox{GravityCam} on a somewhat bigger telescope such as the 3.6m New Technology Telescope (NTT) in La Silla, Chile (Figure 1).  This is an excellent site with median seeing of about $0.75\arcsec$.  From our experience on the NTT we find that 100\% selection yields about $0.3\arcsec$ resolution, and 50\% selection yields better than $\sim\,0.2\arcsec$ (see Figure~\ref{fig:imagesimu}). The NTT has instrument slots in two Naysmith foci with rapid switching between them, so that \mbox{GravityCam} can be installed concurrently with the SoXS instrument \citep{SoXS}.

By removing the tip-tilt components, the phase variance in the wavefront entering the telescope is reduced by a factor of about 7. A theoretical study by \citet{Kaiser} predicts a resulting improvement factor of $\sim 1.7$ on the seeing with 100\% frame selection for a telescope such as the NTT 3.6 m with median seeing of 0.75\arcsec. However, this analysis assumes measuring the mean position of each image, while
experience over many years by users of Lucky Imaging shows that better results are obtained if uses the brightest pixel in each image (usually the centre of the brightest speckle rather than the full image) as a reference position for the shift before addition. By adopting such a procedure, one can achieve a larger improvement factor of $\sim 2.5$ with 100\% frame selection on a 3.6m telescope. With 20\% selection that is increased to at least a factor of 3 and up to a factor of 4 for 1--2\% selection,
as demonstrated by our observational findings. Our "sharper" images are brighter in the core and narrower at their half widths, so that adjacent objects can be separated.
With \mbox{GravityCam} we normally expect to operate with 100\% selection although
the instrument maybe used with smaller percentages to produce higher
resolution images at the cost of reduced efficiency. This performance is
predicted to be possible with significantly less than 100 photons per
frame from the reference star.

It is worth noting that Lucky Imaging is a rather simple technique while more sophisticated approaches have been proposed and tried in practice. Using a fast autoguider that measures the median position of a reference star and computes a correction, which is then used to adjust the telescope guidance, is relatively unsatisfactory because of the nature of the servo loop that carries this out. The atmospheric phase patterns change on short timescales typically tens of milliseconds so that moving the telescope quickly even by a small amount is very difficult. Techniques such as speckle imaging \citep[e.g.][]{Carrano,Loktev} are also difficult to implement, particularly on faint reference stars and over a significant field of view.
With GravityCam, we are constrained by the amount of image processing that may be carried out in real time, which poses limits to the complexity of the adopted technique.

\begin{figure}
\centering
\resizebox{\columnwidth}{!}{\includegraphics{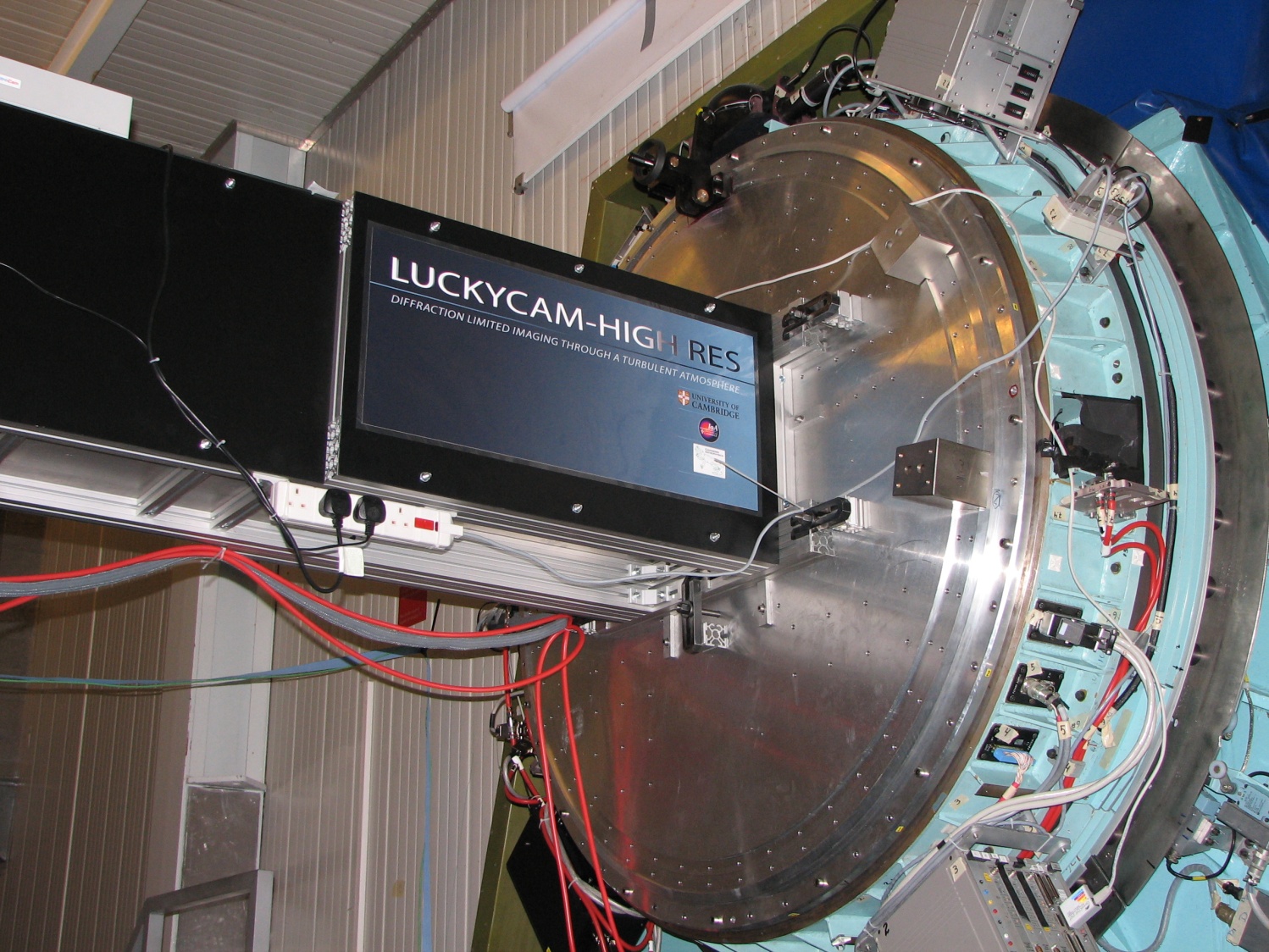}}
 \caption{Prototype of \mbox{GravityCam} detector mounted on one of the Naysmith platforms at the NTT 3.6~m telescope of the European Southern Observatory in La Silla, Chile. 
 This is one example of the instruments used on a number of telescopes to
establish the credentials of the technique on good observing sites such
as La Palma in the Canary Islands and La Silla in Chile. The system
shown here
consisted of a single EMCCD behind a simple crossed to prism atmospheric dispersion corrector (ADC) being run in the standard lucky imaging mode.  There is considerable space to mount an instrument on the telescope which has extremely high optical quality and is located in a top astronomical site.}
\label{fig:proto}
\end{figure}

\begin{figure*}
\centering
\resizebox{\textwidth}{!}{\includegraphics{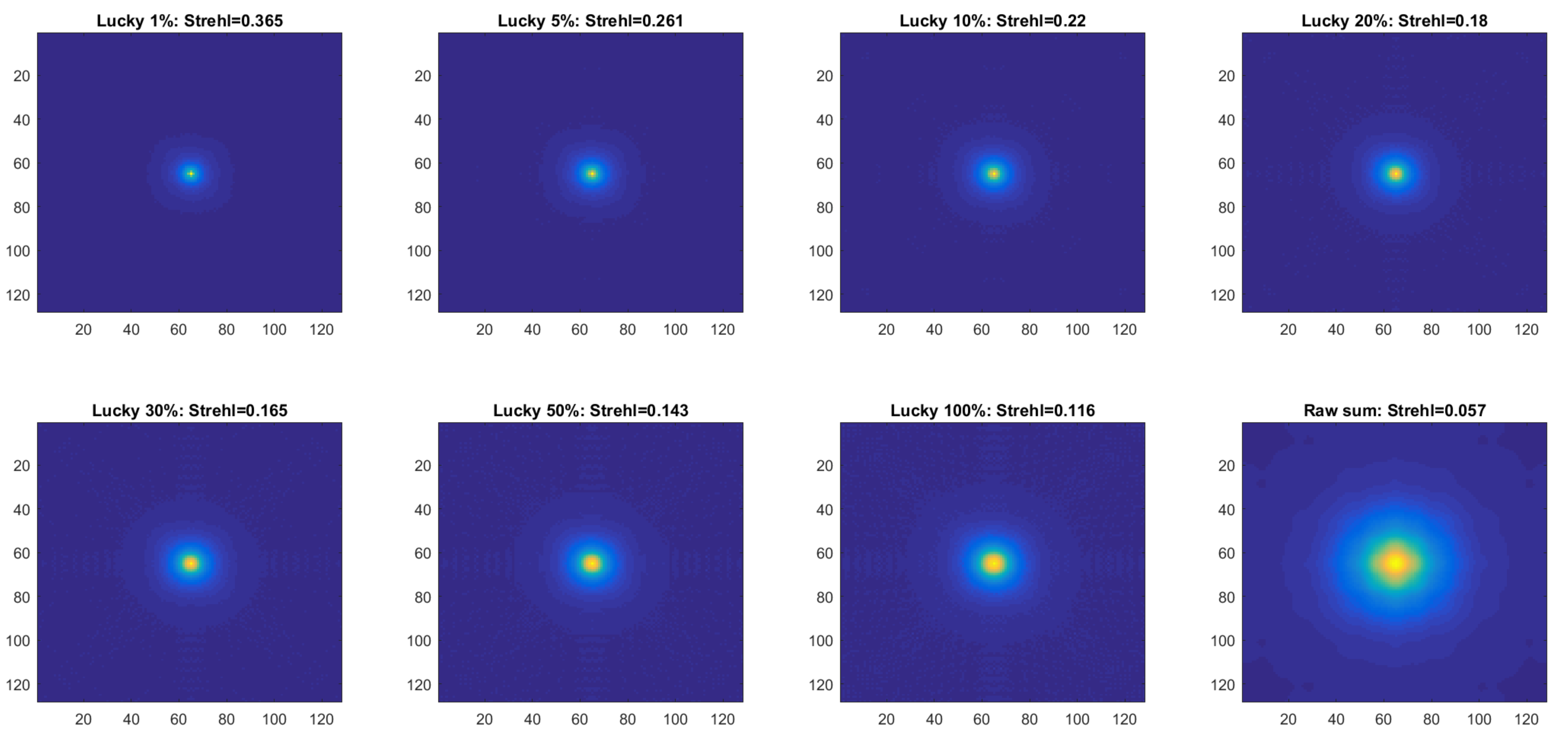}}
\caption{Simulated ESO NTT images of about $3.5\arcsec \times 3.3\arcsec$ size showing the improvement delivered by \mbox{GravityCam} compared with the equivalent raw image with seeing equal to the median value for La Silla of $0.75\arcsec$ FWHM. The images show the result of conventional raw imaging (lower right hand) plus lucky imaging using a variety of selection factors between 1\% and 100\% for image sharpness. The point spread function consists of a narrow core with a faint extended tail. Lucky imaging concentrates light from the halo into the central core. We verified that this simulation for a 2.5m telescope reproduces very closely the results delivered on the NOT telescope on La Palma \protect\citep{Baldwin+2008}.}
\label{fig:imagesimu}
\end{figure*}

\subsection{Technical requirements}

The core requirement for \mbox{GravityCam} is an array of detectors able to run at frame rates $> 10~\mbox{Hz}$ with negligible readout noise.  This will enable very faint targets to be detected.  Until recently such detectors did not exist.  However the development of electron multiplying CCDs (EMCCDs) changed the detector landscape substantially, and were quickly taken up for astronomy \citep{Mackay+2001}. EMCCDs have relatively high readout noise under conventional operation.  However a multiplication register within the EMCCD allows the signal to be amplified before the readout amplifier so that the effective readout noise is substantially reduced in terms of equivalent photons.  The gain may be set high enough to allow photon counting operation.  Even if photon counting is not needed, it is possible to reduce the read noise to a level that is acceptable.  Even more recently the development of high-performance CMOS devices has moved forward very rapidly.  These devices are capable of very low read-out noise levels ($\sim\,$1~electron RMS) while running at fast frame rates (10--30~Hz).  For practical purposes CMOS devices have many of the excellent characteristics of CCDs such as high quantum efficiency, good cosmetic quality, and high and linear signal capacity.  They are also capable of being butted together allowing a large fraction of the area of the field of view of the telescope to be used.  In comparison, EMCCDs typically have only about 1/6 of the detector package area sensitive to light. Although EMCCDs can be packed closely the overall light gathering capability is fairly limited because of the structures needed for a high-speed CCD operation.

Mounted to the 3.6m NTT, GravityCam could cover 
$0.2~\mbox{deg}^2$ in 6 pointings with EMCCDs or $0.17~\mbox{deg}^2$ in a single pointing with CMOS devices.

Ray tracing of the NTT focal plane indicates field curvature with a radius of curvature of 1900~mm. While this has no effect on axis, if left uncorrected the induced defocus at the field edge would lead to a degradation of image quality (80\% encircled energy) to $1.5\arcsec$ radius. This can be partially corrected either by use of a curved/stepped focal plane or a simple single element field corrector to $\sim\,0.6\arcsec$ radius.  To correct the residual field edge aberrations to the lucky imaging limit a more complex corrector will be required \citep{Wynne1968}.

Another factor to be taken into account is atmospheric dispersion. For example at airmass 2.0 ($60^\circ$ Zenith distance) this affect creates an image spread of $\sim\,0.3\arcsec$
between 7000 and 8000~\AA{} \citep{Fillipenko1982}. An atmospheric dispersion corrector \citep{WW86} will therefore also be needed.

In concept, GravityCam is very simple: It is a wide-field imager using conventional silicon imaging detectors. By using Lucky Imaging, we can achieve an improvement in angular resolution by a factor of 2.5--3 from
the resolution that a conventional long-exposure imaging system would give on the same telescope. How to achieve this good resolution, how to achieve it over a wide field of view, and what photometric precision can be achieved, however requires careful thought, further informed by detailed simulations. 

Another instrument, AOLI (Adaptive Optics Lucky Imager) has been under development at Cambridge and the IAC in Tenerife that removes higher-order turbulence terms to give better improvements to image quality 
\citep{AOLI}. A paper is currently in production that describes the much more complicated techniques needed for that instrument, but the same simulation package can be used to predict and optimise the performance  
of \mbox{GravirtyCam} more accurately. We are also fortunate to be able to compare our significant observational experience of Lucky Imaging on a range of telescopes with diameters from 2.5-5 m with the outputs of
the simulation package, showing generally very good agreement.


\subsection{Wide-Field Lucky Imaging}

All the published work on Lucky Imaging has described studies that extend over a very limited field of view typically $< 1\arcmin$ in diameter. In contrast, the field-of-view of the NTT is approximately $30\arcmin$ in
diameter.
Unfortunately, the further a particular target is from the reference star the poorer the image quality will be. This is quantified by the isoplanatic patch size, defined as the diameter within which the image Strehl ratios are reduced by a factor less than 1/e. Both our simulations and observations from the many Lucky Imaging campaigns indicate an isoplanatic patch size of $\sim 1\arcmin$ diameter if we are to achieve  the highest resolution with small selection percentages. Moreover, it is likely to be substantially larger for larger selection percentages.

Given the high stellar density for the vast majority of the fields monitored for our observing programmes (and in fact the crowding being one of the drivers for high angular resolution),
we are not expected to need reference stars as far apart as $1\arcmin$. 
We can therefore process the images over much smaller areas, typically $30\arcsec \times 30\arcsec$, with overlapping adjacent areas for cross-referencing purposes.
The moments of excellent seeing would often be different from square to square. Within each square, the different percentages are accumulated. In this way, the Lucky Imaging performance that we have already demonstrated on many occasions may be achieved with GravityCam. It turns out that processing the smaller images is much easier and quicker than working with images many times that size.

It is also worth noting that we do not need to have a single bright reference object in each field. It is enough to use the cross correlation between the already accumulated images and the image just taken.
In sparsely filled fields the photometric accuracy is maintained. Each frame is added but the offsets derived from the
reference object will be inaccurate. As it is only the tip-tilt correction that is in error, the PSF's are simply smooth out versions of the central PSF. This allows them to be corrected for more easily if
precision absolute photometry is required. 


\subsection{Photometric Precision with GravityCam}

The Lucky Imaging technique is a procedure for taking a subset of images of identical exposure length from a sequence. The subset are chosen on the basis of their image sharpness. Any individual image is either chosen or not chosen. Each image properly represents the flux from the area being studied. If 100 per cent of the images are accumulated, then the summed photon flux will be identical to that which would have been recorded using a conventional long exposure technique. If 10 per cent only are selected, then the photon flux will be precisely 1/10 of that which would be recorded if all the images had been used. This is important because it means that the Lucky Imaging technique does not compromise photometric accuracy. In practice, photometric precision depends on the photon flux from a target star together with the accuracy with which the light collecting power of the instrument may be calibrated. Traditionally, only the very best nights for atmospheric clarity would be used for precision photometry. Some of our research programmes such as asteroseismology studies require precisions significantly better than 1 per cent to be useful. Low levels of atmospheric attenuation, for example due to high altitude cirrus which can be very difficult to detect, will ultimately constrain photometric quality. Different phases of the moon change the sky background level, the brightness of the sky in the absence of any stars in that field. GravityCam will be used for long-term photometric studies by returning
again and again to the same target field. Very quickly the system will establish a very accurate knowledge of the integrated flux from any particular region of the target field. Slight variations will be found because of atmospheric opacity but when we are trying to measure the brightness of one single target object in such a field we must always remember that there are many other targets in the field which we can be certain are, on average, unchanging. The detected image from a field may be corrected to bring the instantaneously detected frame into photometric alignment with the accumulated frames. GravityCam offers a system capable of very precise relative photometry. For none of the studies we propose is absolute photometry required, we simply seek to measure very small changes in the brightness of our target stars.
The photon detection rate with a target star with $I \sim 22.0$ and a broad filter band could be as much as $\sim 500$ photons per second or $\sim$1.8~million photons per hour. It is well established that such photon statistics would lead to better than 0.1 per cent photometric accuracy for conventional CCDs. Moreover, from fits to microlensing light curves obtained with an EMCCD camera at the Danish 1.54m at ESO La Silla \citep{Skottfelt+2015} as part of the MiNDSTEp campaign \citep{MiNDSTEp} since 2009, we found that the photometry follows the light curve as accurately as that from a conventional CCD. 
With the enhanced angular resolution of GravityCam, the photometric accuracy will be better in crowded fields given that we can resolve that part of the sky background that comes from stars of low luminosity.
Therefore, the prospect of achieving photometric precision close to that predicted simply by photon statistics or at least within a factor of two of that is realistic, ultimately limited only by the increasing effect of sky brightness caused by yet fainter target stars eventually making them undetectable. We also established photometric stability with EMCCDs over two-year time-scales, enabling variability studies over such periods \citep{Skottfelt+2015a}. In 2019, we will extensively test a CMOS chip in a camera installed at the Danish 1.54m on these properties.

\section{Planet demographics down to Lunar mass through gravitational microlensing}
\label{Sect:Microlensing}

\subsection{Assembling the demographics}

\begin{figure}
\centering
\resizebox{\columnwidth}{!}{\includegraphics{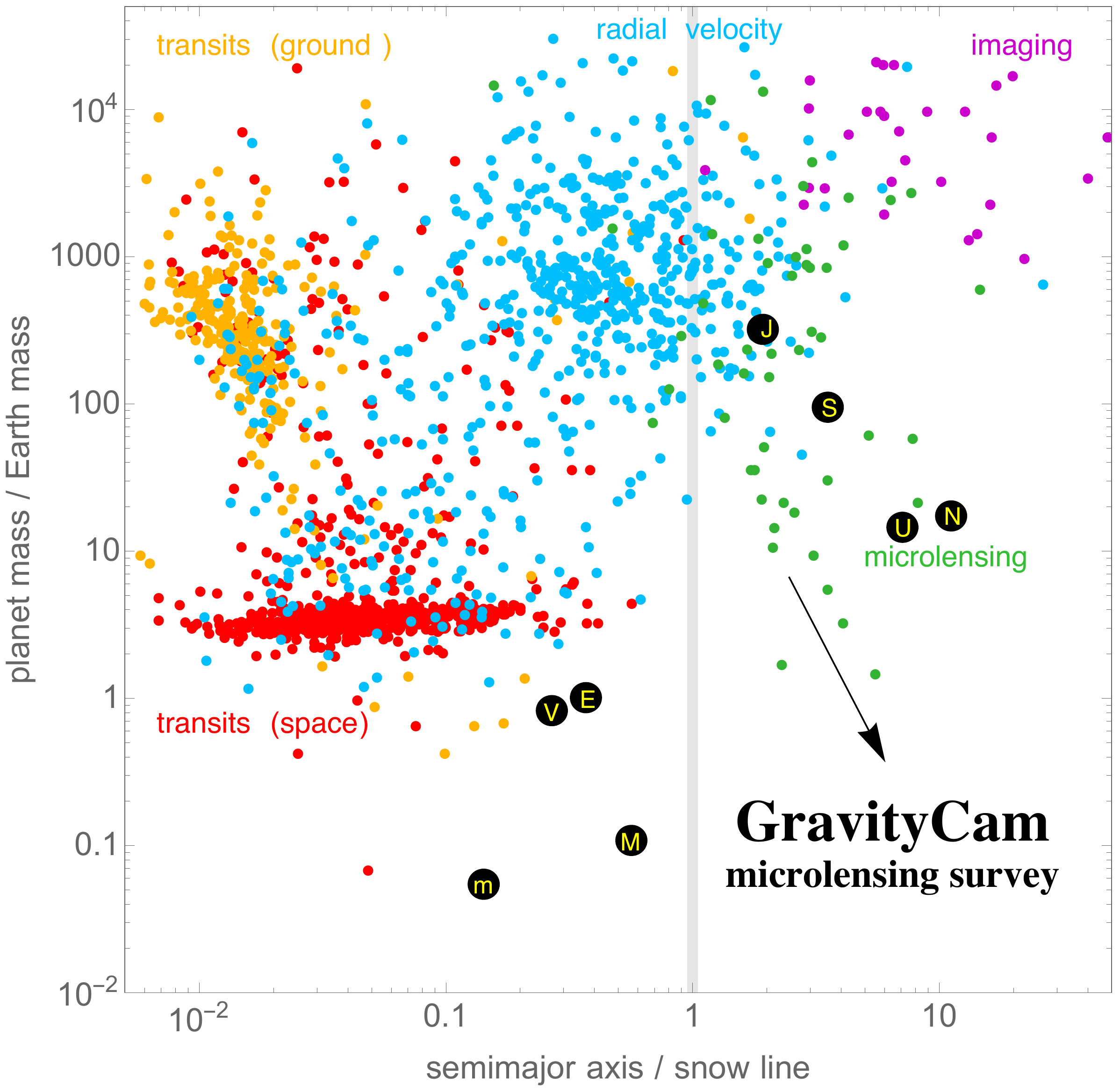}}
\protect\caption{Reported planets by detection technique \citep{exoplanet:encyclopedia} as function of mass and orbital separation relative to the snow line, beyond which volatile compounds condense into solid ice grains. Gravitational microlensing is particularly well suited for exploring the regime of cool low-mass planets. A ground-based survey with \mbox{GravityCam} on the ESO NTT will break into hitherto uncharted territory beyond the snow line and down to below Lunar mass. With $M_\star$ denoting the mass of the planet's host star, the position of the snow line has been assumed to be $a_\mathrm{snow} = 2.7~\mbox{AU}\;(M_\star/M_\odot)$, while the masses $m_\mathrm{p}$ of transiting planets for which only a radius $R_\mathrm{p}$ has been measured has been assumed to be $m_\mathrm{p}/M_\oplus = 2.7\;(R_\mathrm{p}/R_\oplus)^{1/3}$ \citep{Wolfgang+2016}. The planets of the Solar System are indicated by letters m-V-E-M-J-S-U-N.}
\label{fig:techniques}
\end{figure}

While the first planet orbiting a star other than the Sun was only discovered about 20 years ago \citep{MayQel1995}, several thousand planets have now been reported.  It has been estimated that the Milky Way could host as many as hundreds of billions \citep{Cassan+2012} of planets. As illustrated in Figure~\ref{fig:techniques}, the planet parameter space has not been covered uniformly by the various efforts which rely on different techniques, given their specific sensitivities. A comprehensive picture of the planet abundance, essential for gaining proper insight into the formation of planetary systems and the place of the Solar System, can however only arise from exploiting their complementarity. Within foreseeable time, gravitational microlensing \citep{Einstein1936,Pac86} remains the only approach suitable to obtain population statistics of cool low-mass planets throughout the Milky Way, orbiting Galactic disk or bulge stars (two populations with notably different metallicity distributions).

\footnotetext{Source: http://exoplanet.eu, 19 Jun 2017}

\mbox{GravityCam} will overcome the fundamental limitation resulting from the blurring of astronomical images acquired with ground-based telescopes by the turbulence of the Earth's atmosphere, and therefore be competitive with space-based surveys. A \mbox{GravityCam} microlensing survey could 1) explore uncharted territory of planet and satellite population demographics: beyond the snow line and down to below Lunar mass, 2) provide a statistically well-defined sample of planets orbiting stars in the Galactic disk or bulge across the Milky Way, 3) detect substantially more cool super-Earths than known so far, 4) obtain a first indication of the abundance of cool sub-Earths. These results would provide unique constraints to models of planet formation and evolution.

The gravitational microlensing effect is characterised by the transient brightening of an observed star due to the gravitational bending of its light by another star that happens to pass in the foreground. This leads to a symmetric achromatic characteristic light curve, whose duration is an indicator of the mass of the deflector. Gravitational microlensing is a quite rare transient phenomenon, with just about one in a million stars in the Galactic bulge being magnified by more than 30\% at any given time \citep{KirPac1994}. Therefore, surveys need to observe millions of stars in order to find a substantial number of microlensing events, which last about a month. 

\subsection{The GravityCam microlensing survey}
\label{Sec:MicroLensingSurvey}

\begin{figure*}
\begin{center}
\includegraphics[width=4cm]{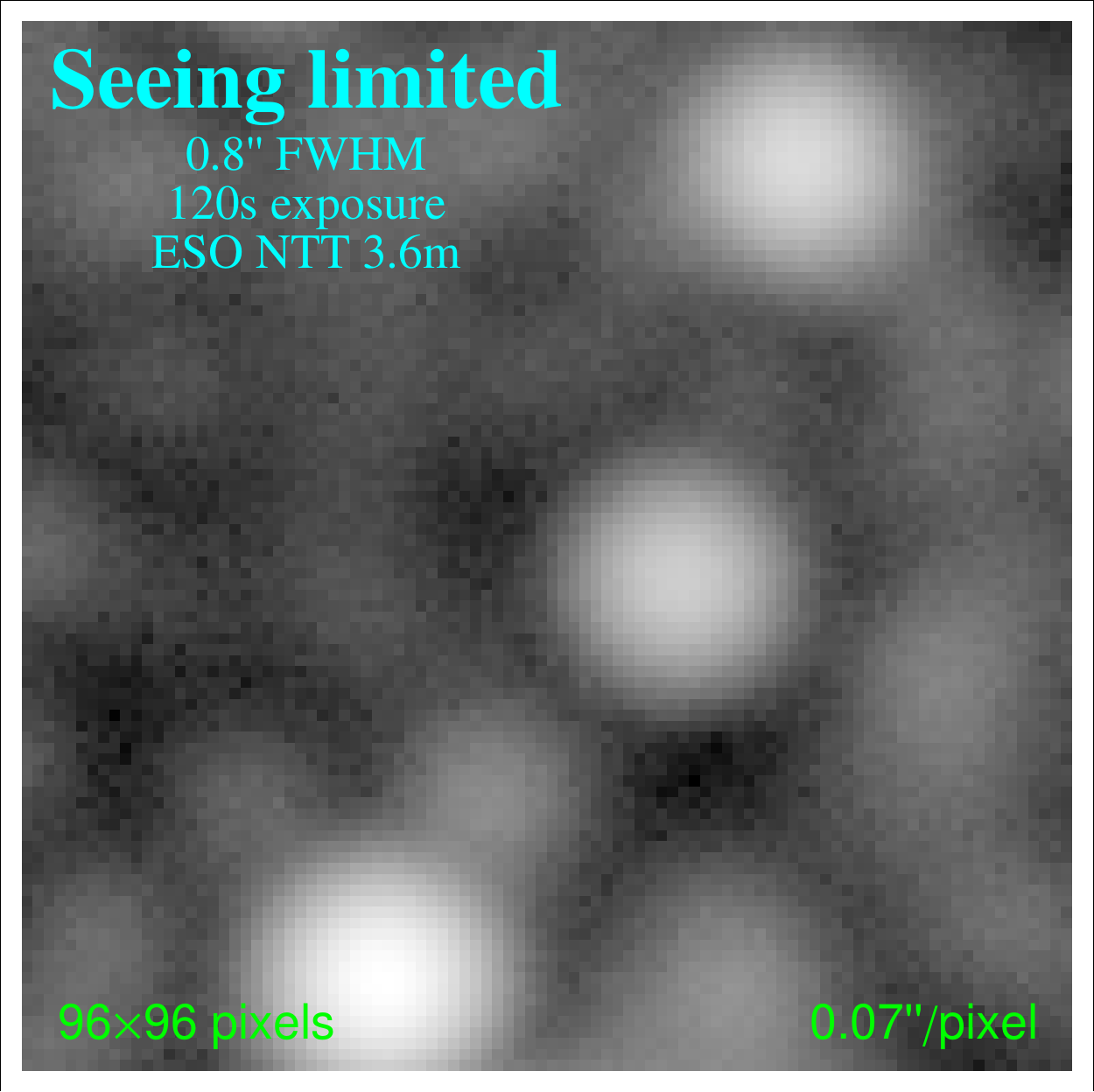}
\includegraphics[width=4cm]{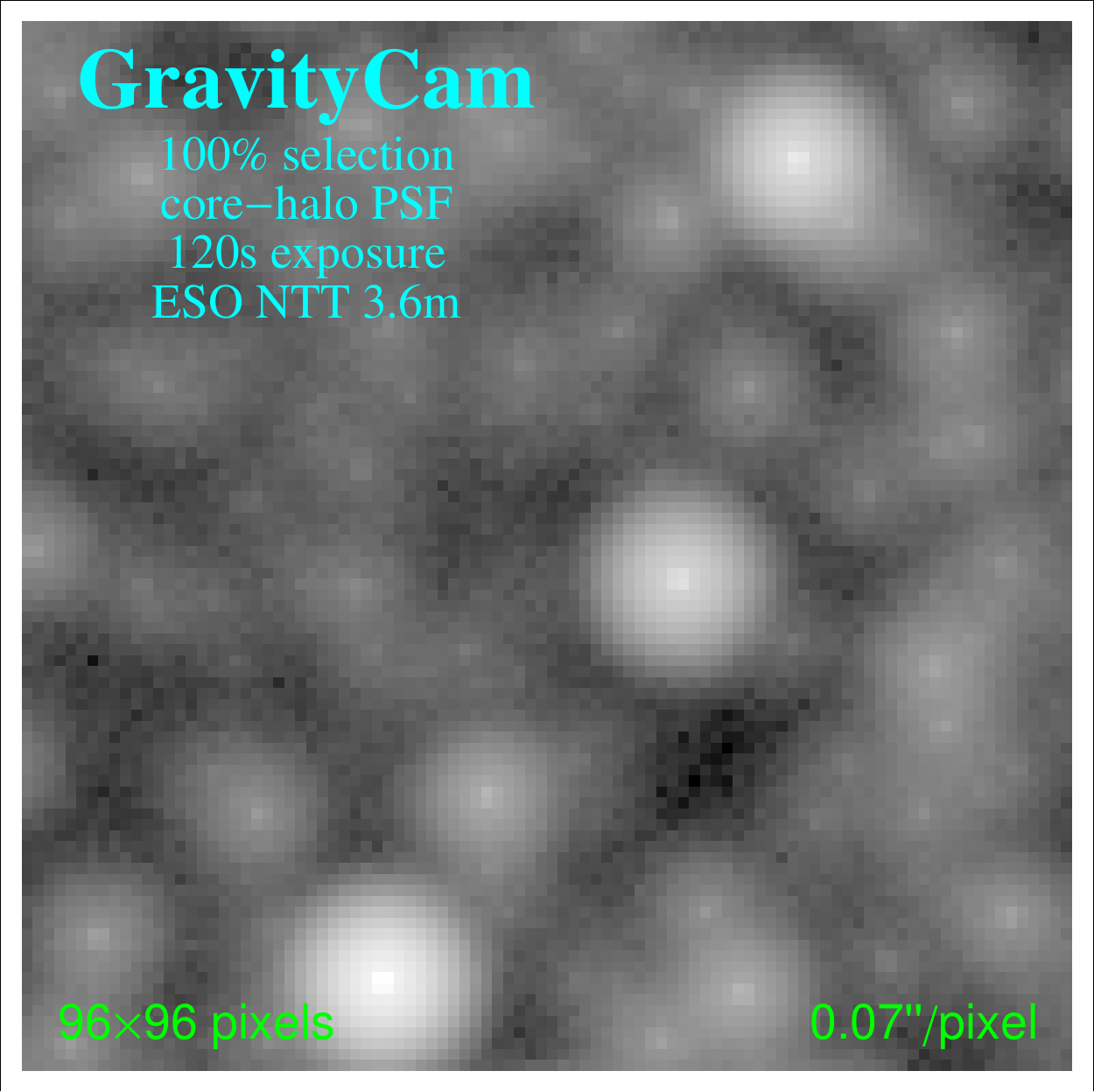}
\hspace*{0.3cm}
\includegraphics[width=4cm]{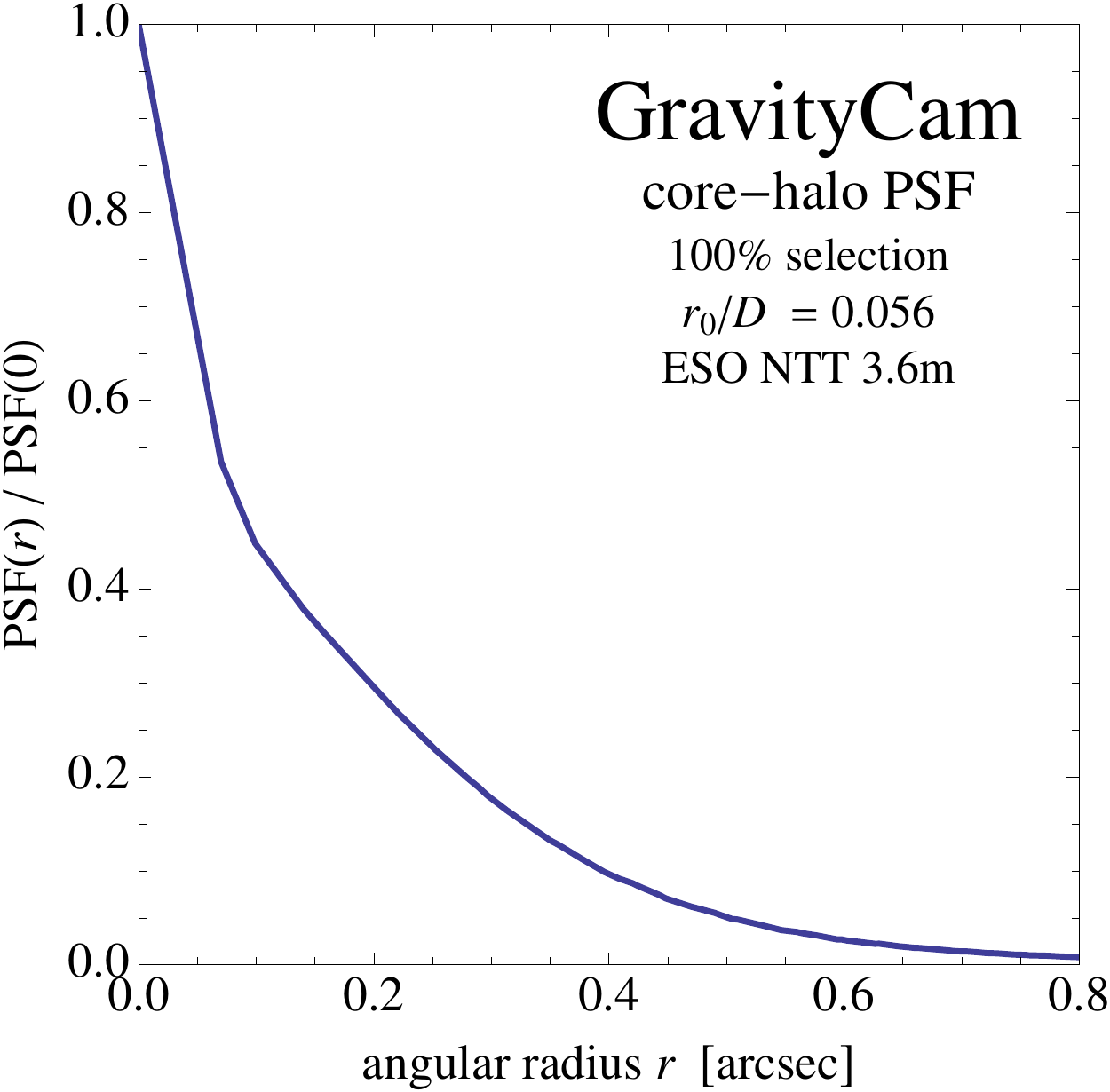}
\end{center}
\caption{Simulated ESO NTT images of about $7\arcsec \times 7\arcsec$ size for 2~min exposures, showing the improvement resulting from \mbox{GravityCam} as compared to being limited by an average $0.8\arcsec$~FWHM, where the core-halo point spread function for 100\% frame selection shown on the right has been adopted, with GravityCam giving $0.07\arcsec$/pixel. }
\label{fig:imagecrowded}
\end{figure*}

The optimal choice of survey fields arises from a compromise between the number of stars in the field, the crowding, and the extinction. Given that most of the Galactic bulge is heavily obscured by dust, the extinction for optical wavelengths can reach levels that make the vast majority of stars practically invisible. However, there are a few ``windows'' with relatively low extinction, the largest of these ``Baade's window'' with a width of about 1$^\circ$, centred at galactic coordinates $(l,b) = (1^\circ,-3.9^\circ)$. Because these fields are very crowded, the high angular resolution achieved with \mbox{GravityCam} will make a crucial difference by dramatically increasing the number of faint (and thereby small) resolved stars in the field as illustrated by the simulated images shown in Figure~\ref{fig:imagecrowded}. Even with 100\% selection of the incoming images, star images will be typically well separated from their nearest neighbour at $I \sim 22$.

\begin{figure*}
\centering
\begin{minipage}{\columnwidth}
\centering
\resizebox{6cm}{!}{\includegraphics{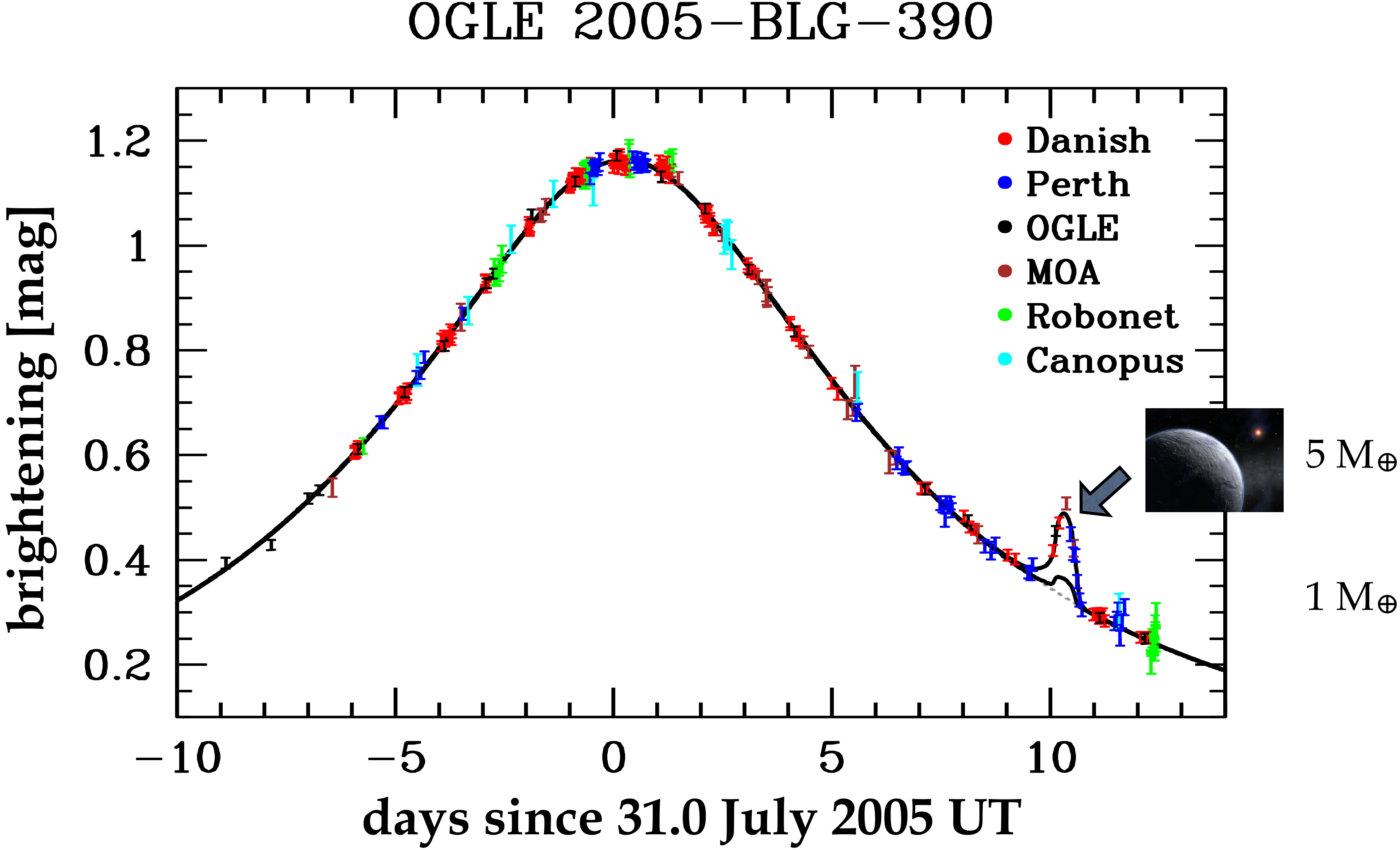}}
\end{minipage}
\hfill
\begin{minipage}{\columnwidth}
\centering
\resizebox{6cm}{!}{\includegraphics{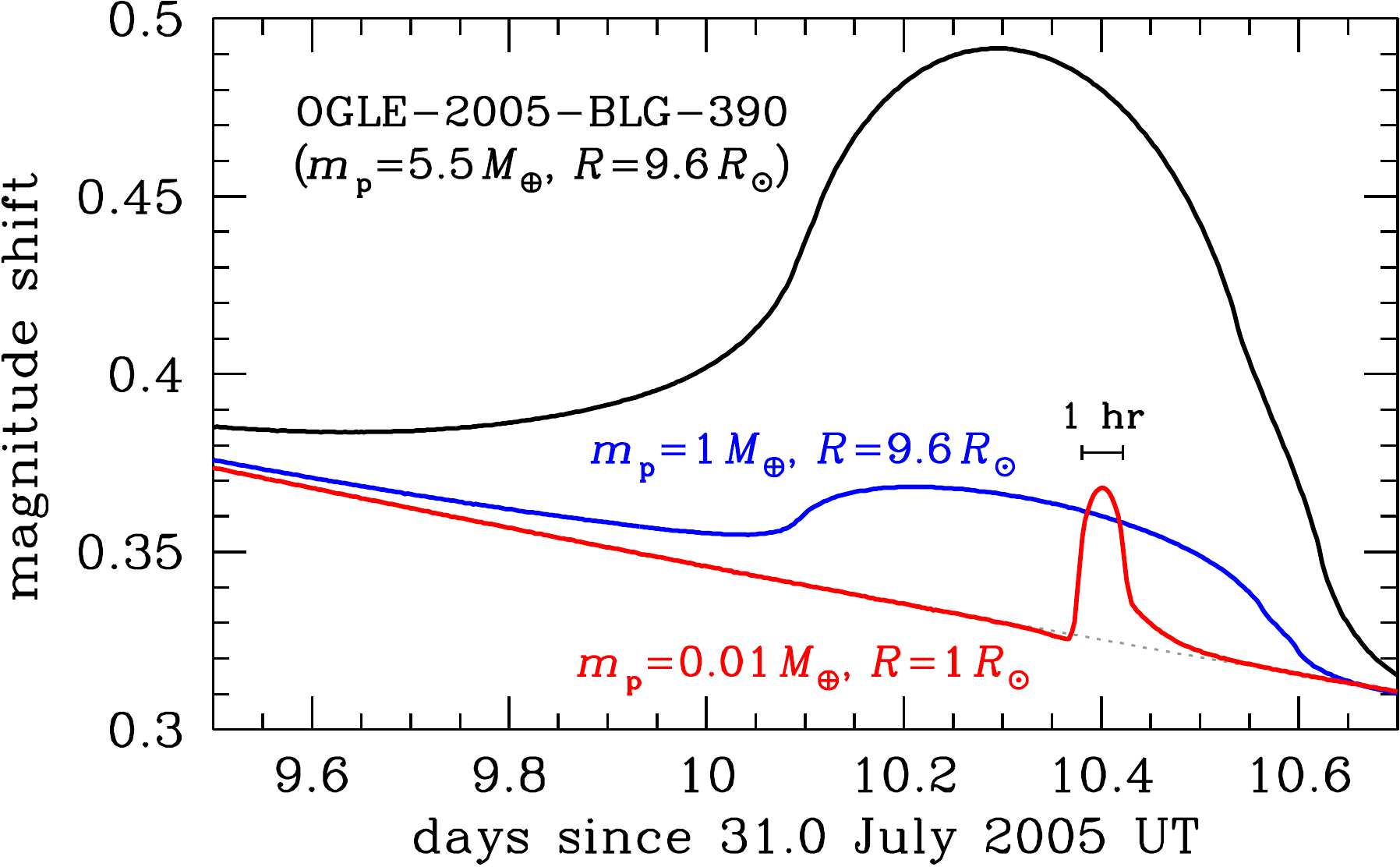}}
\end{minipage}
\caption{({\em left}) Model light curve and data acquired with 6 different telescopes of microlensing event OGLE-2005-BLG-390, showing the small blip that revealed planet OGLE-2005-BLG-390Lb \citep{Bea+2006} with about 5~Earth masses. An Earth-mass planet in the same spot would have led to a 3\% deviation. ({\em right})  Signature of planet OGLE-2005-BLG-390Lb with $m_\mathrm{p} = 5.5~M_\oplus$  and a source star with $R = 9.6~R_\odot$ (black), together with those for an Earth-mass planet in the same spot (blue), and a Lunar-mass body with a Sun-like star (red). Even the latter would be detectable with 2\% photometry and 15~min cadence.}
\label{fig:microsignal}
\end{figure*}

As illustrated in Figure~\ref{fig:microsignal}, a planet orbiting the foreground (`lens') star may reveal its presence by causing a perturbation to the otherwise symmetric light curve \citep{MaoPac1991,GouLoe1992}. Its signature lasts between days for Jupiter-mass planets down to hours for planets of Earth mass or below. Shorter signals do not arise because of the finite angular size of the source star, whose motion relative to the foreground `lens' star limits the signal amplitude by smearing out the effect that would arise for a point-like source star \citep{BenRhi1996,Dominik2010}. Extending the sensitivity to less massive planets therefore means to go for smaller (and thereby fainter) source stars \citep{BenRhi2002}. While cool super-Earths remain detectable in microlensing events on giant source stars ($R \sim 10~R_\odot$) \citep{Bea+2006}, high-quality (few per cent) photometry on main-sequence stars ($R \sim 1~R_\odot$) enables reaching down to even Lunar mass \citep{Pac1996,Dominik+2007}.

\begin{figure}
\centering
\resizebox{\columnwidth}{!}{\includegraphics{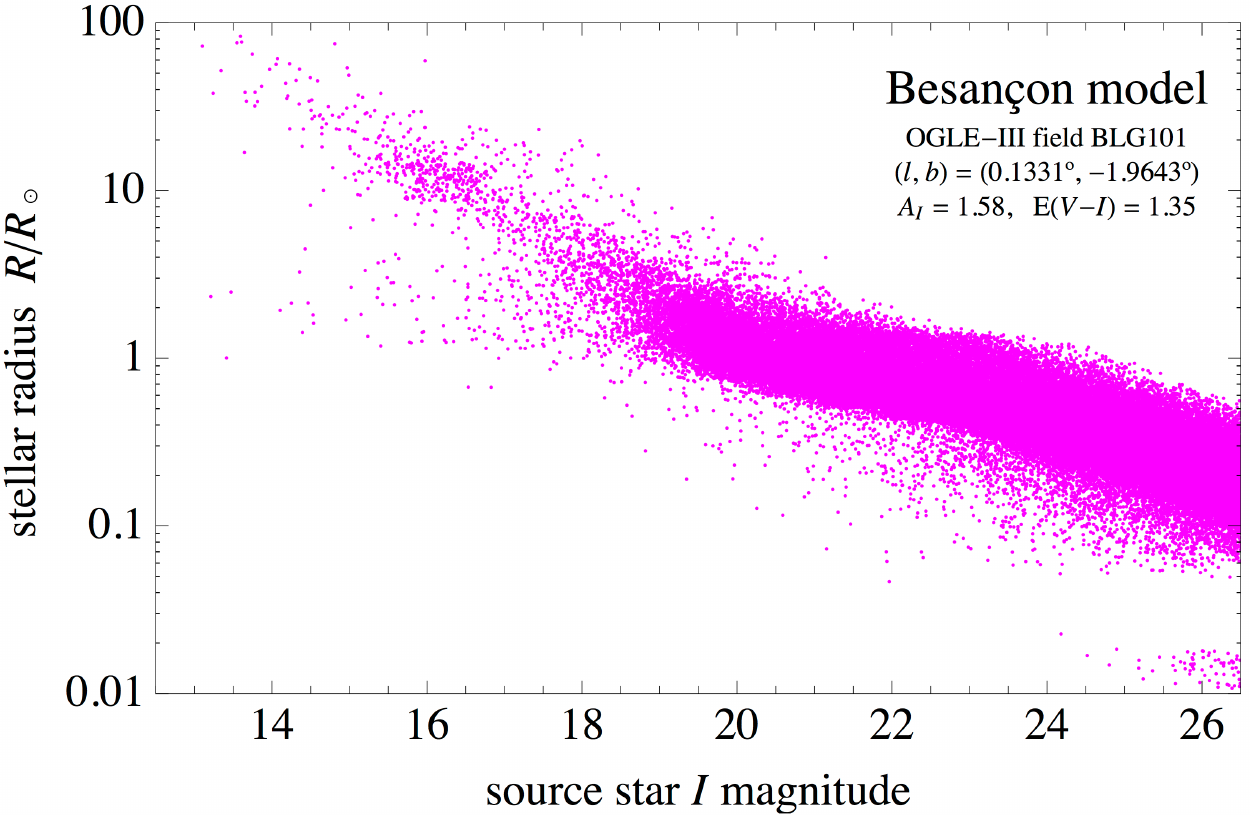}}
\caption{Stellar radius versus $I$ magnitude for stars in the direction of the Galactic bulge, resulting from a Besan\c{c}on population synthesis model \citep{Robin+2003} simulation for the OGLE-III BLG101 field (which has the highest event rate), as well as the extinction and reddening measured from OGLE-III \citep{Nataf+2013}.}
\label{fig:Besancon}
\end{figure}

\begin{figure*}
\centering
\resizebox{12cm}{!}{\includegraphics{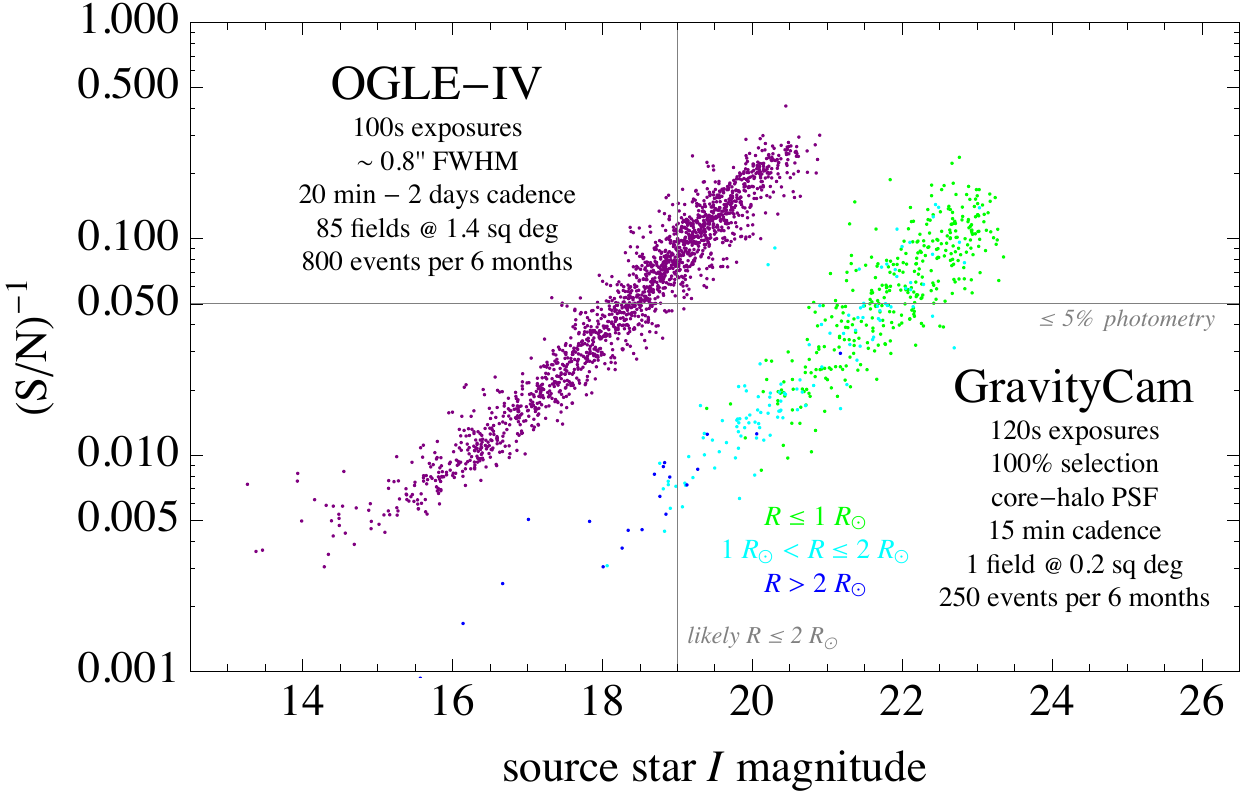}}
\caption{Comparison of performance between OGLE-IV and a microlensing survey with \mbox{GravityCam} on the ESO NTT using EMCCD detectors for resolved stars in the observed fields. With an exposure time of 2~min (similar to OGLE-IV), a single field of $0.2~\mbox{deg}^2$ can be monitored at 15~min cadence. While OGLE-IV misses out on providing $\leq 5$\% photometry on main-sequence source stars, small variations in the brightness of such small stars can be well-monitored with \mbox{GravityCam}. Using CMOS detectors with GravityCam would boost the planet yield by a factor of at least $\sim\,10$, with the area monitored per pointing being 5 times as large and the photometric limits shifting by 0.8~\mbox{mag}.}
\label{fig:OGLE_vs_GravityCam}
\end{figure*}

For those fields in the Galactic bulge most favourable to gravitational microlensing, giant stars start branching off the main sequence at about $I  \sim 19$, with a Solar analogue at 8.5~kpc being at $I \sim 20.3$ \citep{Robin+2003,Nataf+2013}, as shown in Figure~\ref{fig:Besancon}. Current microlensing surveys (such as OGLE-IV; http://ogle.astrouw.edu.pl) use small telescopes (1.3--1.8~m in diameter) and are most fundamentally limited by the typical seeing of 0.75$\arcsec$ FWHM. As Figure~\ref{fig:OGLE_vs_GravityCam} illustrates, with \mbox{GravityCam} on a 4m-class telescope, we can go about 4 magnitudes deeper than OGLE-IV for the same signal-to-noise ratio and exposure time of 2~min, achieving $\le 5$\% photometry for the full range $19 < I < 22$. Most spectacularly, with stars at $I \sim 16$ being about 10 times larger than stars at $I \sim 20$, we go further down in planet mass by a factor 100 at the same sensitivity.

In a single field of $0.2~\mbox{deg}^2$, we can monitor $\sim\,1.0 \times 10^7$ resolved stars with 2~min exposures. With an event rate of $\sim\,5 \times 10^{-5}$ per star and year \citep{Sumi+2013}, we expect $\sim\,$250~events over a campaign period of 6~months. A design with CMOS detectors would increase the field of view for a single pointing by a factor $\sim\,5$, and with a higher effective magnitude limit (by $\sim 0.8~\mbox{mag}$), 
we would expect to gain a total factor $\sim\,10$ in planet yield.
  Given that \mbox{GravityCam} provides the opportunity to infer the planetary mass function for a hitherto uncharted region, the detection yield is unknown, and prior optimisation of the survey strategy is not that straightforward. The choice of survey area and exposure time determines the survey cadence, the photometric uncertainty as function of target magnitude, the number of resolved stars monitored, and ultimately the planet detection efficiency as function of planet mass. Increasing the exposure time at cost of a lower cadence would lead to losing short planetary signatures (unless further follow-up facilities can complement the survey), but the microlensing event rate would increase with more fainter stars being detected. Sticking to a single pointing would provide the opportunity to construct both effective short exposures for high cadence as well as effective long exposures for monitoring fainter objects and obtaining higher angular resolution by means of lucky imaging.

For monitored stars of Solar radius, we hit a sensitivity limit to companions around the foreground `lens' star at about Lunar mass, which means that not only putative planets of such mass could be detected, but satellites as well. Until this happens, the detection efficiency scales with the square-root of the planet mass. If the mass function of cool planets follows the suggested steep increase towards lower masses $\mathrm{d}N/\mathrm{d}[\lg(m_\mathrm{p}/M_\oplus)] \propto (m_\mathrm{p}/M_\oplus)^{-\beta}$ \citep{Cassan+2012}, where $\beta \ge 0.5$, we would therefore detect comparable numbers of planets for each of the mass ranges 1--10~$M_\oplus$, 0.1--1~$M_\oplus$, and 0.01--0.1~$M_\oplus$. The distribution of the detected planets (or the lack of detections) will constrain the slope of the mass function.
 
Given that space telescopes are unaffected by the image blurring due to the Earth's atmosphere, a case for a microlensing survey for exoplanet detection has been made both for ESA's Euclid mission (as potential 'legacy' science) and for NASA's WFIRST (Wide-Field Infrared Survey Telescope)\footnote{http://wfirst.gsfc.nasa.gov}, as one of the competing priorities shaping its design. Euclid is currently planned to be launched in 2020, whereas WFIRST is still at a very early stage of definition with a projected launch towards the end of the 2020s. Euclid provides a 0.55 deg$^2$ FOV with a 1.2m mirror, whereas WFIRST would provide a 0.28 deg$^2$ FOV with a 2.4m mirror, as compared to a 0.17 deg$^2$ FOV with a 3.6m mirror for GravityCam with CMOS detectors on the ESO NTT. The microlensing campaigns with the space telescopes would be restricted to observing windows lasting one or two months only, substantially reducing the planet detection capabilities, given that the median time-scale of microlensing events is around a month \citep{Penny+2013,WFIRST:micro}.

\subsection{Crowded-field photometry with GravityCam}

Charge-coupled devices have been used for many years and their characteristics as detectors are well established. There are features which are just becoming better appreciated but again the quality of photometric work that is being done already is exceptional. The methods for photometry in relatively uncrowded fields are well known having been developed on many telescopes and at many observatories throughout the  world. When thinking about crowded field photometry we have to distinguish between relative photometry which is particularly important for microlensing studies in the bulge of the Galaxy, and absolute photometry.

GravityCam will offer much better angular resolution than is usually available from ground based studies. However we must recognise that in the crowded fields GravityCam will target there are stars at the position of virtually every single pixel on the detectors. Attempting to measure the light from one star is immediately complicated by the contribution from the other nearby stars which may or may not be brighter than our target star. As with any astronomical observation seeking high precision, the effects of variable opacity due to high level cloud or low-level moisture/fog can be significant. Seeing variability and sky brightness variability further complicate matters.

Relative photometry is very much easier in crowded fields because we can be confident that the integrated light across a large patch of the sky will be precisely constant, and any integrated flux variability can be 
immediately calibrated and corrected for. The lucky imaging process will select images not on the basis of percentage but on the basis of an actual achieved resolution per frame. Frames at the same resolution will be combined so that the influence of the point spread function may be better understood. Absolute photometry in crowded fields is more complicated due to source confusion in all-sky photometric catalogues and distance from isolated standard star fields. Taking calibrations at different Zenith distances has to be done with great care and only under the best conditions. 

There is substantial experience in making photometric observations in crowded fields with EMCCDs. For example, the Danish 1.54~m telescope at ESO's La Silla Observatory has played a key role in the follow-up monitoring of gravitational microlensing events since 2003, having provided in particular the crucial data for identifying the then most Earth-like extra-solar planet OGLE-2015-BLG-390Lb \citep{Bea+2006}. In 2009, the telescope was upgraded with a multi-colour EMCCD camera \citep{Skottfelt+2015}. \citet{Harp+2012} made the first investigations of how to optimise the pipeline in order to obtain optimal photometric accuracy, and found RMS of the order 1\,\% from test observations of the core of Omega Cen for stars where scintillation noise dominated the noise budget (magnitudes $< 17$) increasing to a few per cent when photon and excess noise and background crowding dominated the budget. The photometric scatter in the crowded fields were reduced substantially \citep{Skottfelt+2013,Skottfelt+2015a,Figuera+2016} by selecting the very best resolution images (the so-called ``lucky images'', or the sharpest 1\% of the images covering those fraction of seconds where the atmospheric turbulence above the telescope happens to be at minimum) as reference images for the reduction of the rest of the images, and using image subtraction \citep{Bramich2008} instead of Daophot point-spread function reduction. For example, the photometric RMS of EMCCD observations of the relatively bright OGLE-2015-BLG-0966Lb \citep{Street+2016} is well below 1\,\% for 2~min exposure sequences on the Danish 1.54m. 

Our knowledge of the photometric credentials of CMOS detectors is much less well established. A program is currently underway at the Open University to calibrate devices which are very similar to those we 
are likely to use for GravityCam in order to check repeatability, stability, quantum efficiency and uniformity, calibration issues etc. So far the results are very promising. Indeed we would not expect very great differences from the point of view of imaging devices as they are based on using the same silicon structures used in CCDs. There is also substantial experience in using infrared detectors such as those made by 
Teledyne and Rockwell which use CMOS readout electronics with mercury cadmium telluride detector elements bump bonded to the CMOS components. We know from those devices that there are indeed problems for example with residual charge (reading out a pixel does not completely empty it) and this is something which will be the subject of future investigation. However there is a great deal of knowledge about CMOS structures and how these are used because of the ubiquity of CMOS detectors in devices such as mobile phones and handheld digital cameras which are required to achieve extraordinarily high imaging quality \citep{Janesick2014}.

\section{A unique database for optical variability}

\subsection{Stellar variability and asteroseismology}

Given that many stars belong to classes which are known to be variable, studies of stellar variability are a collateral benefit of virtually any optical survey. Gravitational microlensing surveys with \mbox{GravityCam} described above can produce precision relative photometry on as many as 90 million stars with each measured for extended periods every clear night over an entire observing season. With a limiting magnitude $I \sim 27$ for 1~h exposures and the high angular resolution, the sample obtained with \mbox{GravityCam} extends much further towards fainter stars than OGLE-IV, while the cadence is much higher than for LSST. 

Moreover, virtually every star will show low levels of variability simply because of the complex structure within each star that reflects the internal structure of the star and how soundwaves within the star propagate.  Helioseismology studies of the Sun have used such data and have given a great deal of information about the internal structure of the Sun \citep{Gough2012}.  While photometric studies of more distant objects cannot resolve the surface as is possible with the sun,  they can still provide substantial information about the internal structure of the star.  A recent example of the methods and results that may be obtained from the studies is given by \citet{Bowman+2016}.

Such asteroseismology studies critically rely on precision photometric measurements of the star over a long period of time. The high stellar density in fields towards the Galactic bulge enables such high
precision given that in each and every frame the photometric calibration is provided by comparing the target star with the mean flux from all the others.  This suppresses very effectively any variations in atmospheric transmission from atmospheric haze or thin cloud cover.

At a photometric accuracy better than 500~$\umu$mag in one hour, {\mbox GravityCam} is expected to provide data on around 5 million stars with $I < 19.5$ each night over a period of 6 months with \mbox{EMCCDs}, or 25 million stars with $I < 20.3$ with CMOS chips. Brighter stars will be imaged to even higher photometric accuracy as this accuracy is simply determined by the relative precision with which the baseline photometry from the field is established.  Over the critical timescales for stellar oscillations of 1--100~hours very high accuracy data will be generated.  While it had previously been believed that it was extremely difficult to extract the full frequency spectrum of the oscillations within a star if the data available had significant gaps as is inevitable with the single ground-based instrument, new methods have been developed that get round this problem, very much in the way that other disciplines have had to cope with missing data or incomplete sampling \citep{Pires+2015}. 
 
The observing cadence of \mbox{GravityCam} is also well-suited for detecting and monitoring eclipsing binaries.

\subsection{Sub-second variability from accretion onto compact objects}

Massive stars end their lives as compact objects, i.e. neutron stars and black holes, leading to systems in which a compact object accretes material from a companion star in a binary orbit. The short dynamical times make rapid variability a defining characteristic of such accreting compact binary systems, known as \lq X-ray binaries\rq, given that such variability {\em in X-rays} is well documented (e.g.\ \citealt{bellonihasinger90}, \citealt{vanderklis95} and many others). 

\begin{figure}
\centering
\resizebox{\columnwidth}{!}{\includegraphics[angle=90]{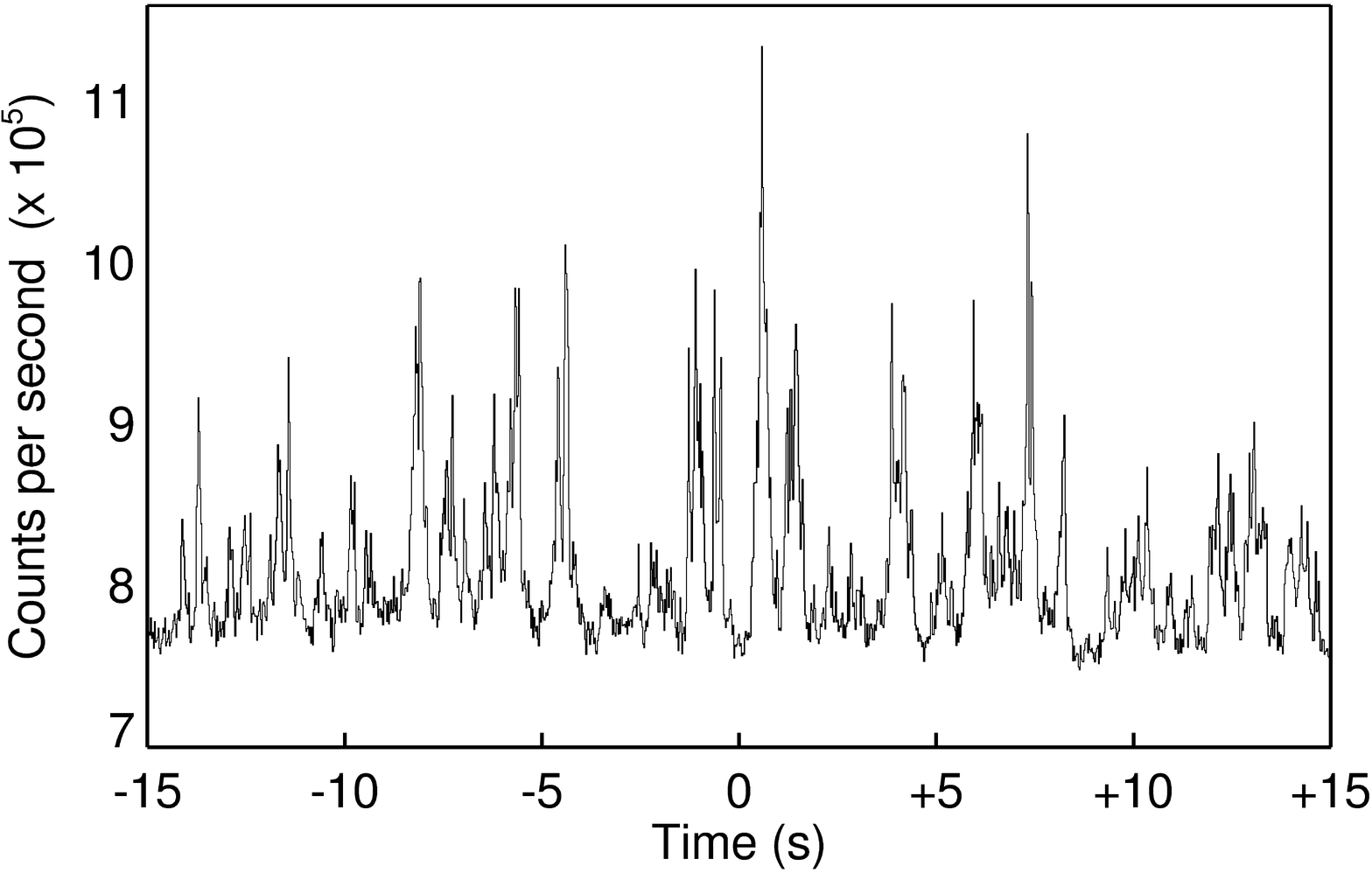}}
\caption{Example of rapid optical variability from an accreting black hole binary V404\,Cyg. The figure shows a short 30\,s segment of an ULTRACAM $r'$ light curve from 2015 June\,26. Fast sub-second flares were visible throughout these observations, with complex structure of the flares visible on $\sim$\,100\,ms timescales and shorter. These sub-second flares are interpreted as non-thermal synchrotron emission from the base of the relativistic jet in this source. The full data set is described by \citet{gandhi16}.}
\label{fig:rapid_variability}
\end{figure}

However, few studies exist on rapid optical variations of such objects. Optical photons are typically generated as thermal radiation from viscous stresses in the outer (cooler) regions of the accretion disc. Alternatively, X-rays from a central hot electron \lq corona\rq\ can irradiate the outer regions and be reprocessed to the optical regime. Both pathways provide a way to map the physical and dynamical state in the outer parts of accreting flows on timescales of order $\sim$\,1--10\,s (e.g.\,\citealt{obrien02}), and at least some observations have shown the presence of other mechanisms at work on fast timescales (see \citealt{uttleycasella14} for a review). This argues against irradiated components such as the outer disc where the fluctuations are expected to have longer characteristic times. The best evidence for such behaviour exists for XTE\,J1118+480 \citep{kanbach01}, GX\,339--4 \citep{motch82, gandhi10}, Swift\,J1753.5--0127 \citep{durant08}, and V404\,Cyg \citep{gandhi16}. An example light curve segment is shown in Figure\,\ref{fig:rapid_variability}. Similar evidence also exists in some neutron star binaries \citep{durant11}, though on somewhat longer characteristic timescales. In at least two cases (GX\,339--4 and V404\,Cyg), the fastest variations have a red spectrum, further arguing against thermal reprocessing which is expected to show blue colours.

By cross-correlating the fast optical fluctuations with strictly simultaneous X-ray observations, time delays between the bands have been mapped, revealing a complex mix of components. The shortest delays span the range of $\sim$\,0.1--0.5\,s with the optical following the X-rays, interpreted as the propagation lag between infalling (accreting) material and outflowing plasma from the base of the relativistic jet seen in these systems \citep{kanbach01,malzac04,gandhi10}. Knowing the location of the jet base synchrotron emission is critical for constraining jet acceleration and collimation models (e.g.\,\citealt{markoff05}) and the optical delays appear to be constraining these to size scales of order 10$^3$ Schwarzschild radii, though this remains to be tested in detail. 

On slightly longer timescales of order $\sim$\,1--5\,s, there is evidence of the optical variability preceding the X-rays in anti-phase \citep{kanbach01,durant08,gandhi10,pahari17}, often interpreted as synchrotron self-Compton emission from the geometrically-thick and optically-thin corona lying within the disc \citep{hynes03,yuan05, gandhi10, veledina11}. Such a medium may also undergo Lense-Thirring precession, resulting in quasi-periodic oscillations (QPOs) in the optical and X-ray light curves which give insight on the coronal dynamics \citep{hynes03,ingram09,gandhi10}. 

Fast optical timing observations can thus provide quantitative and novel constraints on the origin and geometry of emission components very close to the black hole cores. But with only a handful of observations thus far (the 4 objects mentioned above), this field remains at its incipient stages with much degeneracy between the models cited above. 
Progress has been hindered, in large part, due to the lack of wide availability of fast timing instruments with low deadtime. This is now starting to be addressed with instruments such as ULTRACAM \citep{ultracam} on the NTT and ULTRASPEC \citep{ultraspec} on the Thai National Telescope which are capable of rapid optical observations, though neither instrument is available throughout the year. SALT is also approaching optimal observing efficiency (following recent mirror alignment corrections), and has a fast imager SALTICAM capable of rapid optical studies \citep{salticam}, but the telescope pointing is constrained so most objects are visible only for short periods during any given night. A few other specialised optical instruments also exist. But another hurdle has been the lack of sensitive X-ray timing missions for coordination with optical timing. This is also now changed with the launch of the {\em AstroSat} and {\em NICER} missions. 

GravityCam can play a major role in this emerging field. Its natural advantages include fast sampling capabilities on timescales of $\sim$\,0.1\,s, low deadtime, and a wide field of view allowing simultaneous comparison star observations. Typical peak magnitudes of black hole X-ray binaries are $V$\,$\sim$\,15--17 (Vega), which should be well within reach from a 4\,m class telescope. Time tagging of frames with GPS is a possibility that can enable such science. 



\subsection{Transits of hot planets around cool stars}

A further type of variability that will show in the \mbox{GravityCam} data sets is the dip in light from a star produced by a planet passing in front of it \citep{Struve:exoplanet}.  Normally the confirmation of the discovery of a planet requires that its transit is recorded several times.  The depth of the dip in the light from the star and the length of the transit gives important information about the planet and its orbital parameters. Unlike gravitational microlensing, which favours the detection of cool planets, such planetary transits favour the detection of hot planets, given that planets at larger distances will have larger orbital periods, and a larger orbit also makes it less probable that the planet is accurately aligned with its host star to pass in front.

While planet population statistics depend on the properties of the respective host stars, the faintness of M stars for optical wavelengths makes these more difficult targets in surveys than FGK stars. However, M stars are more favourable to detecting small planets due to their smaller radii.

The characteristics of \mbox{GravityCam} differ substantially from other instruments, which will lead to a quite different sample. 
NGTS \citep[Next Generation Transit survey;][]{2017PASP..129b5002M}, and TESS  \citep{TESS} use small telescopes with low angular resolution (NGTS: 20cm, 5\arcsec/pixel; TESS: 10.5cm, 21\arcsec/pixel), restricting these surveys to bright nearby stars. For MEarth \citep{2008PASP..120..317N}, the pixel scale is smaller (0.76\arcsec/pixel), but the telescope diameter is small as well (40 cm).
In contrast, \mbox{GravityCam} can deliver angular resolutions $\sim 0.15\arcsec$ with a 3.6m telescope, and with CMOS detectors would reach $S/N \sim 400$ 
at $I \sim 18$ with $2~\mbox{min}$ of integration. \mbox{GravityCam} thereby addresses the crowding of Galactic bulge fields and enables photometry with 
$1$--$10~\mbox{mmag}$ precision that is needed to reliably detect exoplanet 
transits.

Using data from the VVV  \citep[VISTA Variables in Via  L\'actea,][]{2010NewA...15..433M}  survey, a preliminary investigation  \citep{2014A&A...571A..36R} estimates about 15,000 objects per square degree with
 $12< K_\mathrm{S} < 16$ and colours consistent with M4--M9 dwarfs, as well as  $\sim 900$ such objects per square degree for $K_\mathrm{S} < 13$. These M dwarfs are located relatively nearby at $0.3$--$1.2~\mbox{kpc}$ and their near-infrared colours are only moderately affected by extinction. With this magnitude range corresponding to about $17< I < 21$ for early/mid-M stars \citep{2013ApJS..208....9P}, one therefore finds $\sim\,3000$ potential M dwarfs (or $\sim\,180$ brighter M dwarfs with $I < 18$) for a single $0.2~\mbox{deg}^2$ GravityCam field.

GravityCam moreover provides a high time resolution, favourable to studies of transit timing variations \citep[e.g.][]{TTV,2009A&A...507..481C,2014A&A...565A...7C}.

\section{Other applications}
\label{Sect:other}

\subsection{Galactic star clusters}

It is well known that star clusters host numerous tight binaries. They are involved in the dynamic evolution of the globular clusters and may also produce peculiar stars with anomalous colours and/or chemical composition \citep{JHL2015}. It has been also predicted that the first population in the globular clusters, with multiple populations, could contain a higher fraction of close binaries than the second generation \citep{Hong+2016}. Recently, \citet{CarBen2017} explained the extreme horizontal branch stars of the open cluster NGC~6791 introducing tight binaries. These recent results indicate that the detection of tight binaries, with periods between approximately 0.2 up to several days, could be important in the interpretation of some crucial observational issues of open and globular star clusters.

GravityCam is not only well suited to study stellar multiplicity due to the provided high angular resolution, but moreover accurate photometric measurements in crowded fields can provide detection of the eclipse features of the binaries. In addition, high-resolution imaging of clusters can yield proper motion measurements to be used for kinematic studies (e.g. \citealt{HSTPROMO}).
A survey with \mbox{GravityCam} on selected open clusters with anomalous horizontal branches, and in the central regions of massive globular clusters, in particular where multiple populations have been detected, with repeated observations lasting some nights, would therefore be valuable.

The picture of galactic globular clusters that we got so far from HST (e.g. \citealt{ACS}; \citealt{HSTPROMO}) is far from complete, leaving us with many very crowded and reddened clusters at low galactic latitude
as well as some very distant halo clusters that have never observed with HST, or only with a single
epoch, which does not permit inferring proper motions.

GavityCam could also add very important data
on new variable stars (mainly RR Lyrae stars), mostly close to the
relatively little explored cluster central regions. Their accurate
photometry
and improved statistics are fundamental in order remove
He-age-metallicity degeneracies and constrain the He abundance (from
the luminosity level of the variable gap in the Horizontal Branch) which is a recent
hot topic in globular clusters (e.g. \citealt{PopulationsBook}; \citealt{Kerber2018}).

Experience from successful high-resolution monitoring of the central regions of some globular clusters have already been obtained with the small $45\arcsec \times 45\arcsec$ field of the EMCCD camera at the Danish 1.54m telescope at La Silla (e.g. \citealt{Figuera+2016,Figuera2+2016}), leading to the discovery of variable stars previously not identified with HST images. In particular, by avoiding saturated stars in the field of the globular clusters observed, the discovery of variable stars around the top of the red giant branch became possible. This explicitly demonstrates that  globular cluster systems still need further studies and are not as well understood as one might have
thought.

\subsection{Tracing the dark matter in the Universe}

Most of the mass in the Universe is thought to be cold dark matter, forming the skeleton upon which massive luminous structures such as galaxies and clusters of galaxies are assembled. The nature and distribution of dark matter, and how structures have formed and evolved, are themes that are central to cosmology. A prediction of Einstein's general theory of relativity, that mass deflects electromagnetic radiation, underpins the power of gravitational lensing as a unique tool with which to study matter in the Universe. Galaxies and clusters of galaxies can act as gravitational lenses, forming distorted images of distant cosmic sources. If the alignment between the lens and source is close, then multiple, highly magnified images can be formed; this is known as strong lensing. If the alignment is less precise then weakly lensed, single, slightly distorted images that must be studied statistically result. Observations and analysis of these gravitational lensing signatures are used to constrain the distribution of mass in galaxies and clusters, to test and refine our cosmological model and paradigm for structure formation. 


\begin{figure*}
\centering
\begin{minipage}{8.5cm}
\centering
\resizebox{8cm}{!}{\includegraphics{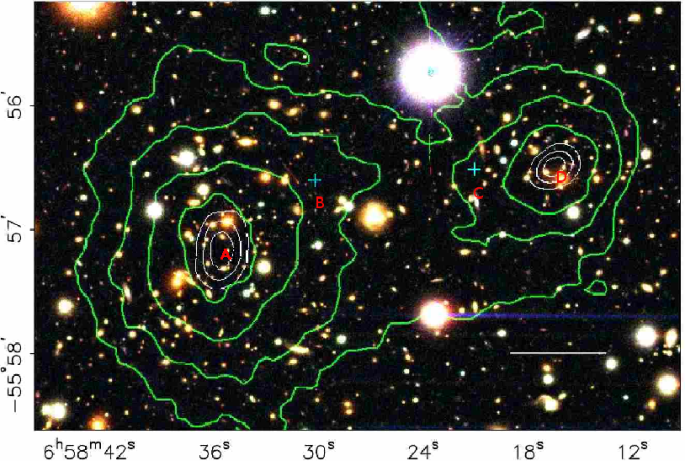}}
\end{minipage}
\hfill
\begin{minipage}{8.5cm}
\centering
\resizebox{8cm}{!}{\includegraphics{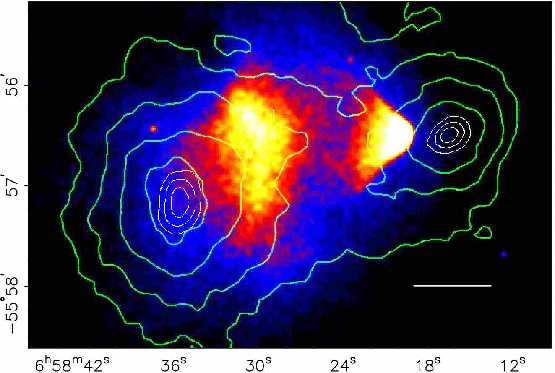}}
\end{minipage}
\caption{The Bullet Cluster; ({\em left}) optical image with contours showing projected mass derived from lensing; ({\em right}): same lensing mass map contours now with X-ray image showing location of hot gas (dominant component of normal matter).  \citet{Clowe+2006} showed that the mass budget is dominated by dark matter.  The projected mass derived from lensing with the HST is excellent but with ground based studies it is extremely hard to recover with any accuracy.}
\label{fig:bullet}
\end{figure*}

High resolution is crucial both for measuring the small distortions of lensed sources in the weak lensing regime, and to study the details of strongly lensed, highly magnified and distorted multiple images. In fact, weak shear studies look for distortions much smaller than the seeing size. As shown by \citet{Massey+2013}, and discussed further by \citet{Cropper+2013}, inherent biases in the measurement of weak lensing observables caused by the size and knowledge of the PSF scale quadratically with the size of the PSF compared to the background galaxy size. The challenge addressed by \mbox{GravityCam} is to obtain exposures for which the PSF is both small in size and has planarity across the field of view.

For the bullet cluster, \citet{Clowe+2006} found the majority of the total matter density being offset from the stellar and luminous X-ray gas matter densities -- and therefore difficult to explain as being associated with non-dark matter mass (Figure~\ref{fig:bullet}). In this system, two clusters have violently impacted and shot through each other, with the hot gas (the dominant luminous matter component) responsible for X-ray emission in both clusters being self-impeded and remaining between the two mass peaks (dominated by dark matter) seen on the lensing mass map (Figure~\ref{fig:bullet}). Both the resolution of such observations and the number density of distant background sources are critical aspects in the accuracy with which mass distributions such as this can be mapped.  

Given a small and stable PSF, the requirements for accurate mass reconstruction are a large number density of objects (to reduce shot noise in the averaged measured ellipticities of the background galaxies), and a wide field of view (for an efficient observing schedule). The resolution and sensitivity of \mbox{GravityCam} would be comparable to the ACS instrument on HST, yet covering a wider field of view, hence excellent for such studies. 

Observational data and N-body simulations suggest that about 10\% of the dark matter in galaxy clusters is in the form of discrete galaxy-scale substructures, the remainder being distributed in a larger-scale dark matter halo. Strong lensing in clusters has been used to investigate the truncation of the dark matter halos of cluster members (e.g. \citealt{Halkola+2007}). 

\begin{figure*}
\begin{center}
\begin{minipage}{6cm}
\centering
\resizebox{6cm}{!}{\includegraphics{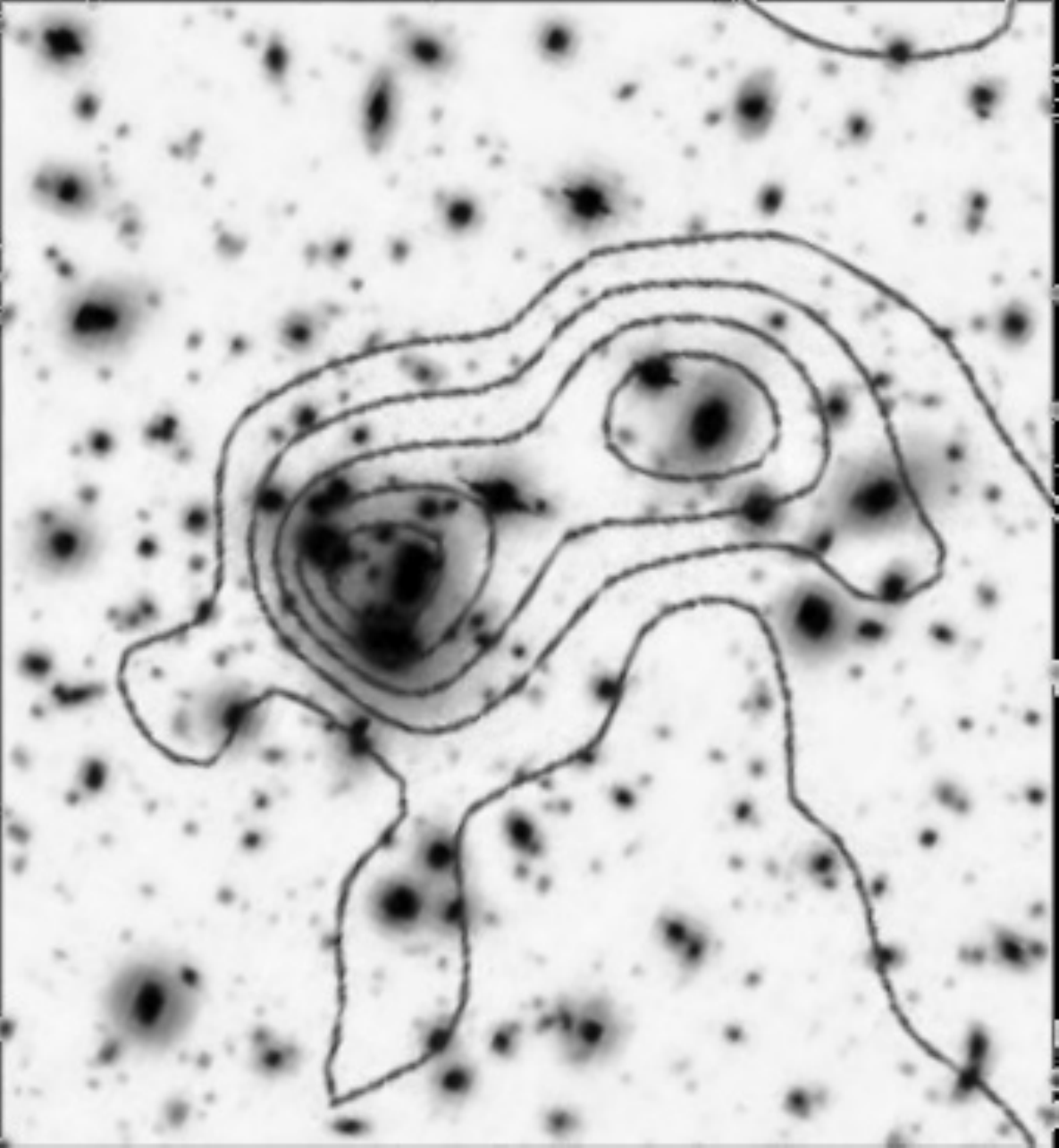}}
\end{minipage}
\hspace*{3mm}
\begin{minipage}{6cm}
\centering
\resizebox{6cm}{!}{\includegraphics{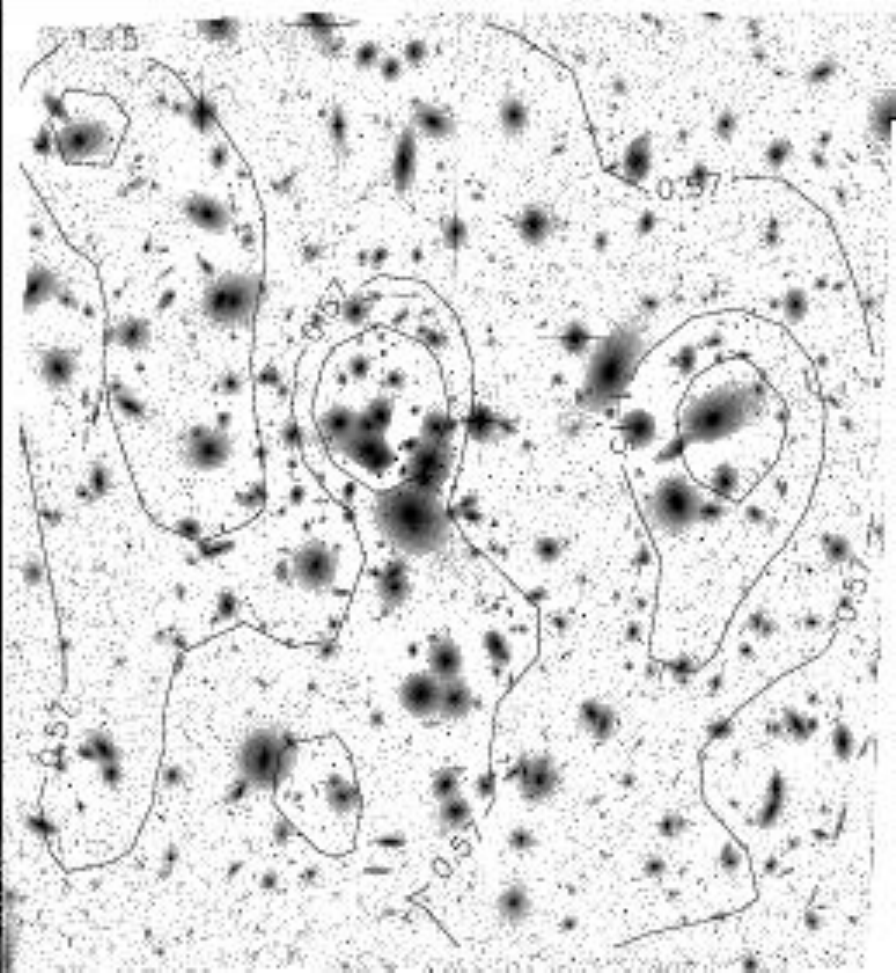}}
\end{minipage}
\end{center}
\caption{Central portion of ground-based Subaru image ({\em left}) and space-based HST/ACS image ({\em right}) of Abell~1689 with contours showing the reconstructed mass distribution from distortion measurements.  Note that space-based data are high-resolution and typically deeper, whereas ground-based data typically cover a larger area (tens of arc minutes compared to a few arc minutes).}
\label{fig:Abell1689}
\end{figure*}

Analysis of the distortion signal has been successfully used to determine the distribution of substructure in the galaxy cluster Abell~1689 using both ground-based, wide-field images from Subaru (Figure~\ref{fig:Abell1689}, left panel, \citealt{Okura+2007}), and using deep HST/ACS images of the central part of the cluster (Figure~\ref{fig:Abell1689}, right panel, \citealt{Leonard+2007}).  \mbox{GravityCam} will combine these characteristics 
by producing deep high-resolution wide-field images.


While \mbox{GravityCam} shares several features with the Euclid spacecraft, scheduled to launch in 2020, there are some characteristics that make these instruments complementary for maximising the return on these science drivers. Euclid uses a 1.2~m diameter telescope with an array of CCD detectors covering a field of view about three times
larger than that of \mbox{GravityCam}.  The image resolution is almost the same as that of \mbox{GravityCam}, while the NTT with its larger mirror has 9 times the collecting area.  As the spacecraft will be located at the L2 position, communication limitations mean that exposure times will be relatively long. Whilst the Euclid cosmic shear survey will benefit from space-based imaging over a wide field of view, there will be unique systematic effects that are required to be modelled, for example charge transfer inefficiency \citep{Holland+1990} and the brighter fatter effect \citep{Downing+2006}. Therefore complementary observations at a similar level of space-based quality will be valuable in testing and confirming the measurements from Euclid. LSST observations will not be of sufficient resolution to provide this complementary data, and therefore GravityCam can play a unique role within the context of the Euclid experiment.
Furthermore, the Euclid weak lensing measurements are made in a single broad-band (RIZ, 500--800~nm), and require space-based narrow band imaging over a smaller area to calibrate so-called ``colour gradient'' effects caused by the broad-band (see e.g. \citealt{Semboloni+2013}). \mbox{GravityCam} will be able to follow up on HST by providing such data over large fields-of-view. 

 \mbox{GravityCam}  also has ideal capabilities for the follow-up of the background sources that are strongly gravitationally lensed by the foreground galaxy clusters, with the whole strong lensing region fitting within a dithered  \mbox{GravityCam}  observation. Euclid is expected to discover $\sim\,5000$ galaxy clusters that have giant lensing arcs \citep{Euclid}. With angular magnifications of up to $\sim\,100$, the stellar populations will be resolvable by \mbox{GravityCam} on scales approaching $25~\mbox{kpc}$, unachievable with the single wide optical passband of Euclid. High magnification events have also been discovered through follow-ups of wide-field infrared and sub-mm surveys or through follow-ups of lensing clusters (e.g. \citealt{Swinbank2010,Iglesias-Groth2017,Canameras2015,DiazSanchez2017}). Future Stage-IV CMB experiments and proposed missions such as CORE will supply many more high magnification events (e.g. \citealt{DeZotti2016}). There are already indications in a small number of objects for the entire far-infrared luminosities to be dominated by a handful of extreme giant molecular clouds (e.g. \citealt{Swinbank2010}). The sub-arcsecond angular scales resolved by \mbox{GravityCam}'s optical imaging, tracing stellar populations, will be well-matched to the atomic and molecular gas and dust traced by ALMA in these objects.

\subsection{Solar System Objects}
\label{Sect:SSO}

  Observations of Solar System Objects (SSO) with GravityCam will take
  advantage of both the improved spatial resolution compared with
  natural seeing and the high time resolution of photometry. The first
  will be of use in studying binary asteroids, for example, which is
  currently done with adaptive-optics camera on 8m class telescopes
  \citep[e.g.][]{Margot-Ast4}. By resolving binary pairs their mutual
  orbits can be measured, allowing derivation of the mass of the system,
  and therefore density, the most fundamental parameter to understand
  the composition and structure of rocky bodies \citep{Carry2012}. There
  are 
  21 such binaries known in the main asteroid belt which
  could be studied, and about 80 (fainter ones) in the Kuiper Belt. As
  with all other areas of astronomy, the improved S/N for point sources
  using the seeing-corrected images from GravityCam will also be of use
  for measuring orbits and light curves (and possibly colours, depending
  on the availability of multiple filters in the final design) of faint
  SSOs. 

\begin{figure}
\begin{center}
   \includegraphics[width=\columnwidth]{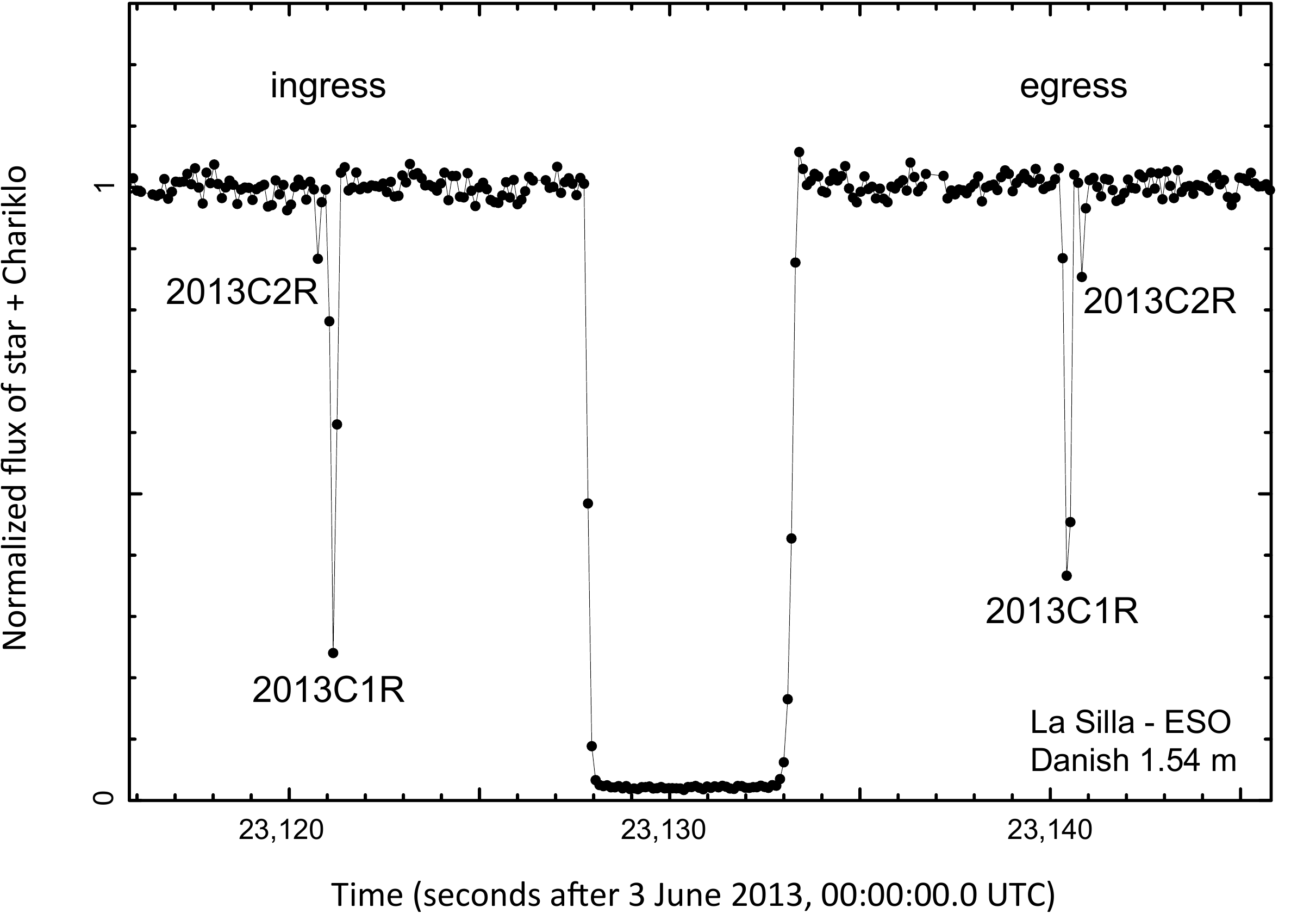} 
\end{center}
   \caption{Discovery of the rings around Chariklo via occultation of a background star. This high speed (10 Hz) photometry was collected at the 1.54m Danish telescope at La Silla with its Lucky Imaging camera.}
   \label{fig:chariklo-lc}
\end{figure}

\begin{figure}
   \centering
   \includegraphics[width=7.6cm]{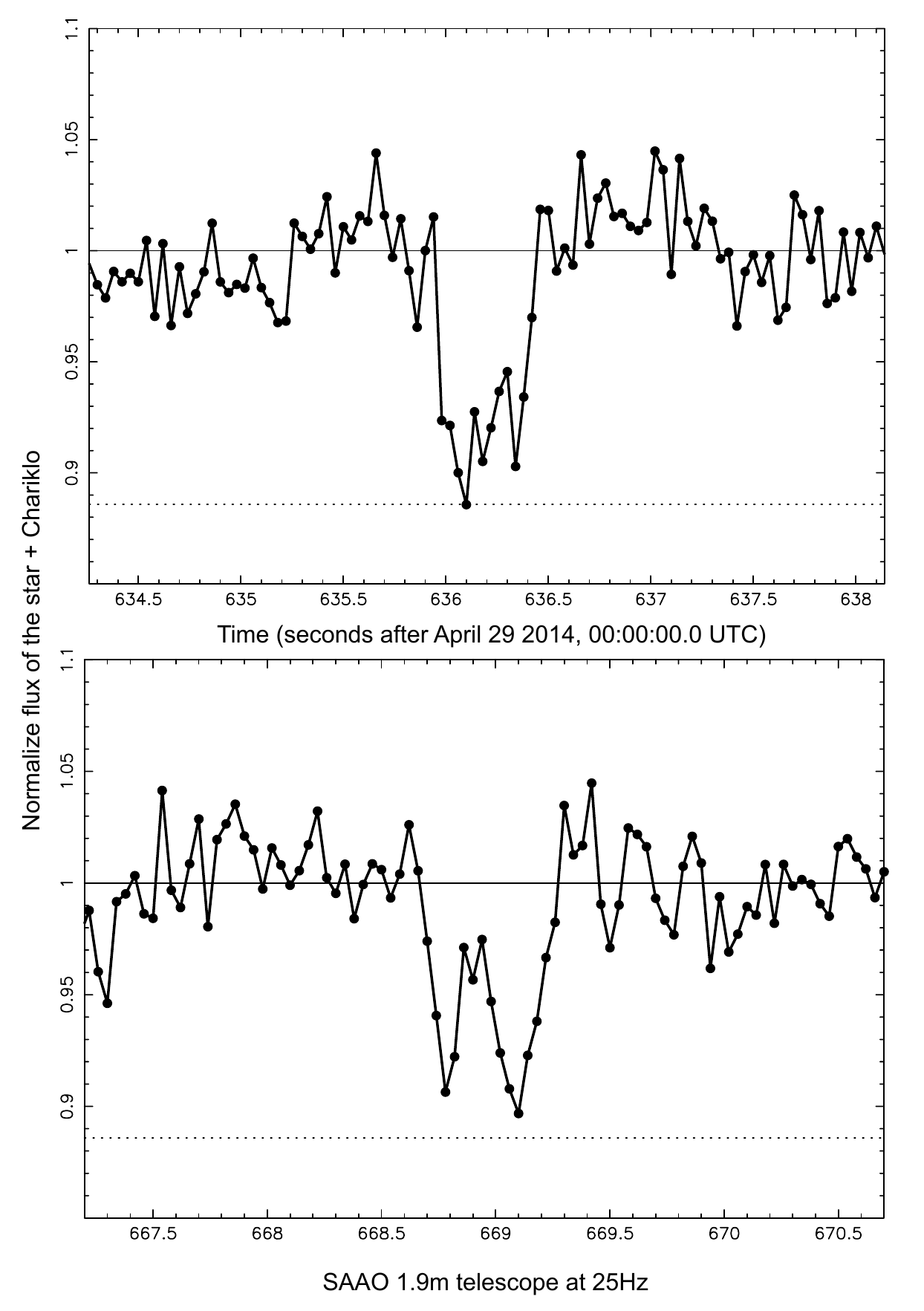} 
   \caption{Occultation of a bright star by Chariklo's rings, taken at even faster frame rate (25~Hz) at the SAAO 1.9m telescope.}
   \label{fig:lc-highspeed}
\end{figure}

  The primary science case for SSOs though is in the time domain. The
  high speed of readout of GravityCam, combined with its large FOV,
  makes it ideal for studying small bodies via occultation of background
  stars. Occultation studies are a powerful way to probe small bodies;
  timing the length of the blink of the background star gives a direct
  measurement of the size of the body, for SSOs too small to directly
  resolve. Multiple chords across the same body, from different
  observatories that see subtly different occultations, can reconstruct
  its shape in a way only rivalled by spacecraft visits
  \citep{Durech-Ast4}. Occultations
  can also probe atmospheres on larger bodies (e.g. on Pluto --
  \citealt{Dias-Oliveira2015,Sicardy2016}) and discover satellites
  \citep{2013-PSS-87-Timerson} or even ring systems
  \citep{Hubbard1986,Chariklo}. For this work high speed photometry is a
  major advantage. In the discovery of the ring system around the
  small solar system object Chariklo, the 10 Hz photometry from the Lucky Imaging camera
  \citep{Skottfelt+2015} on the Danish 1.54m telescope at La Silla was
  critical as it not only showed the drop due to the rings to be deep
  and of short duration (and therefore opaque narrow rings rather than a
  diffuse coma), but even resolved the two separate rings (Figure~\ref{fig:chariklo-lc}). Other
  telescopes with conventional CCD cameras saw only a single partial
  dip, as the occultation by the rings represented only a fraction of
  the few-second integrations. GravityCam would enable target
  of opportunity observations of occultations by known bodies to probe
  size, shape and their surrounding material, with the advantage that
  the larger diameter of the NTT primary mirror, compared with the 1.54m
  Danish telescope, would allow occultations of fainter stars to be
  observed, greatly increasing the number of potentially observable
  alignments. 
  With brighter stars, the high frame rate possible with GravityCam would allow  study of  fine structure. With the highest readout and a typical shadow velocity, we would have sub-kilometre spatial resolution across the rings of Chariklo, allowing measurement of the variation of  optical depth along its width (Figure~\ref{fig:lc-highspeed}), permitting the study of the particle size distribution, and dynamical structures caused by gravity waves and/or oscillation modes caused by Chariklo's mass distribution \citep{Michikoshi2017}.
  
  Occultations are expected to flourish as a technique in
  the next few years, as the accuracy of star positions and minor body
  orbits is vastly improved by Gaia astrometry, meaning that the
  accuracy of event predictions, and therefore the hit rate for
  successful observations, will improve
  \citep{2007-AA-474-Tanga}. 

In addition to such targeted observations, a great strength of GravityCam will be in discovery of unknown minor bodies via occultations caught by chance during other observations. This technique has long been proposed as the best way to detect very small or distant SSOs \citep{Bailey1976,Nihei2007}. This is the only way to discover Oort cloud objects, which would be far too faint to directly detect even with the largest telescopes, and very small objects in the Kuiper Belt, essential for understanding the size distribution down to the size of a typical comet nucleus and below (and therefore constraining models of comet origins, e.g. \citealt{Schlichting2012,Davidsson2016}). The Kuiper Belt size distribution is still poorly understood, even with the latest constraints from counting craters on Pluto from New Horizons images \citep{2015Icar..258..267G}.

The chance of detecting an occultation depends on many factors (e.g. diffraction effects dependent on the size and relative velocity of the SSO, finite source size and colour of the star, filter choice, S/N and time sampling -- see \citealt{Nihei2007}), but in the end mostly depends on how many stars can be monitored, and for how long. 
A productive survey should maximise the number of star-hours observed. Pointing into the ecliptic plane is more likely to discover Kuiper Belt objects (KBOs), but Oort cloud objects could be detected anywhere on the sky.
In fact, the galactic and ecliptic planes overlap, and Baade's window (most suitable for microlensing observations) is by chance also a good place to hunt for SSOs, as it has ecliptic latitude of only $-6.3^\circ$.

When performing the microlensing survey, the combination of the wide field of view with rich star fields will mean that it will be possible to have fast photometry on many stars at any given time, which can be mined for chance occultations by small bodies. 
With EMCCD detectors, 5\% photometry for individual exposures at 10--30~Hz would be possible down to a limiting magnitude of $\sim\,15.5-14.5$, respectively, while the limit would be lower by
$\sim\,0.8~\mbox{mag}$ with CMOS detectors. This gives a fair number of stars per FOV in the galactic bulge fields that can be followed at good S/N for strong constraints on the smallest objects, and it will be complemented by the many more that will be monitored at S/N comparable to CHIMERA \citep{CHIMERA} or TAOS \citep{Alcock2003,Lehner2006}. The brightest stars in the bulge fields are likely to be giants, but a reasonable number of main sequence stars with small apparent diameter, more sensitive to occultation by small bodies, will also be included (see Figure~\ref{fig:Besancon}). Dedicated occultation searches could probably operate with each detector binned to increase sensitivity and/or allow faster readout, but the major advantage for GravityCam is that the occultation search will mostly come for `free', as the data from the microlensing searches (or any other observation) can be mined for events. Between 100~billion (with EMCCD detectors) and 1~trillion (with CMOS detectors) star hours per bulge season at S/N~$\sim$~5--10 should be achievable, with fairly conservative assumptions. The only requirement is that the data processing pipeline also records photometry for sources detected in individual readouts, as well as performing the alignment and stacking to produce the seeing corrected frames for deep imaging.

\section{The GravityCam instrument}
\label{Sect:Instrument}

Large telescopes with wide fields of view are relatively uncommon unless they have been designed with complicated (and often very expensive) corrector optics.  The detectors we propose have pixel sizes in the range of 10--24~$\umu\mbox{m}$. This matches well the NTT.  For the purposes of this paper we will assume that \mbox{GravityCam} will be mounted on the Naysmith focus of the telescope and that the detectors will have $16~\umu\mbox{m}$ (86 milliarcsec) pixels.  The NTT has a $0.5^\circ$ diameter field of view with Ritchey-Chretien optics.  The plate scale is $5.36\arcsec/\mbox{mm}$.  This allows us to mount \mbox{GravityCam} on the NTT without any reimaging optics apart from a field flattener integrated as part of the detector package 
front window. The simplest version of \mbox{GravityCam} consists of a close-packed array of detectors all of which are operating in synchronism to minimise inter-detector interference.  The light from the telescope passes through an atmospheric dispersion corrector (ADC) which is essential to give good quality images free from residual chromatic aberration from the atmosphere.  This is particularly important in crowded field imaging as well as where accurate measurement of the shape of the galaxy is critical.  Telescopes such as the LSST that hope to avoid using an ADC are likely to have significant problems particularly with studies such as the weak shear gravitational lensing program described above.  In front of the detectors may be mounted interchangeable filter units should they be required.  It is worth mentioning that gravitational lensing is a completely achromatic process (while occultations are mostly achromatic) and, unless there is a good scientific case to use a filter, filters may be dispensed with in order to give the highest sensitivity for the survey in question.  However in some cases it will be important to get detailed information of the stars and galaxies being targeted. Asteroseismologists for example need to know the colours in order to understand the internal structure of stars, which will require the availability of appropriate filters for these measurements.

The \mbox{GravityCam} instrument enclosure will be mounted on the image rotator that is part of the NTT Naysmith platform (see Figure 1) and no other mechanisms apart from a filter changer and the ADC are needed. The detector package will need to be cooled to between $-50~^\circ{}$C and $-100~^\circ{}$C to minimise dark current in the detectors.  The detector package will therefore need to be contained within a vacuum dewar enclosure, and using either a recirculating chiller or liquid nitrogen.  Each detector will have its own driver electronics and direct interface with its host computer.  A modular structure is essential to allow individual modules to be replaced quickly in case of module failure so that the entire camera dewar may be vacuum pumped and cooled in good time before the next observing night.  The volume of the entire \mbox{GravityCam} package mounted on the Naysmith platform of the NTT might be about one cubic metre.  The computer system necessary to serve the large number of detectors would be very much bigger, but does not need to be located particularly near to the detector package.

\section{GravityCam detector package}

There are specific requirements on the detectors for \mbox{GravityCam}.  Key to the concept is the need to read the detectors at relatively high frame rates, at least by astronomical standards.  In order to deliver a significant improvement in angular resolution then a minimum detector frame rate is probably around 10~Hz \citep{Baldwin+2001}.  This allows individual images to be checked for image quality and, if different selection percentage subsets are to be combined separately, this quality selection process carried out. Higher frame rates up to perhaps 30~Hz would enable the system to work under a wider range of observing conditions.  The choice of detector frame rate has considerable effect on the processing requirements of the computer system. When fields at high galactic latitude are being imaged, there may be fields that are relatively empty of reference stars with which to judge the quality of each frame.  The higher frame rates will reduce the signal-to-noise on the reference stars and, at those latitudes, that may make it harder to achieve the necessary performance without reducing the frame rate.  However, in most fields a simple cross correlation between the reference frame (rather than a singular specific reference object) and the new frame will allow the tip tilt errors to be eliminated. This is easy and
computationally straightforward. The quality of that cross correlation can also be used to measure the quality of the new frame.
When working with very crowded fields for gravitational microlensing there are many stars that can contribute to these measurements, and it will be possible to find a sufficiently bright standard star in the field, even at high frame rate exposures.  Nevertheless there is a trade-off. 

\begin{figure}
\centering
\resizebox{\columnwidth}{!}{\includegraphics[angle=270]{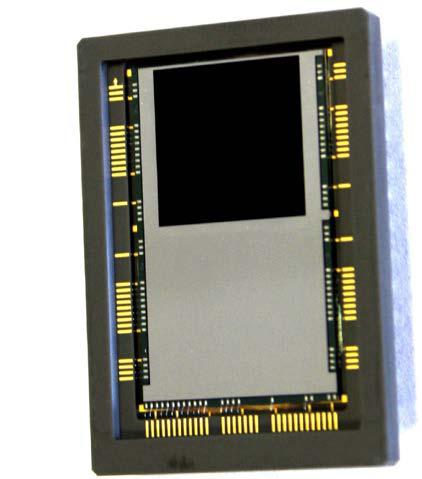}}
 \caption{The CCD201 (Teledyne E2V, UK) is currently the largest area EMCCD capable of working at frame rates above 10 Hz.  It has $1024 \times 1024$ pixels of $13~\umu\mbox{m}$.  The internal gain register allows its operation with essentially zero readout noise at the expense of much reduced full well capacity.  However at fast frame rates this is much less critical. The EMCCD shown here is already available commercially and could be used for GravityCam without any further development.}
\label{fig:EMCCD}
\end{figure}

 One possible choice for the \mbox{GravityCam} detector is an electron multiplying CCD (EMCCD: Figure~\ref{fig:EMCCD}).  These have the advantage of being able to work with internal gain that can produce images with essentially zero readout noise.  The gain necessary to achieve that is typically several hundred and that can reduce the maximum full well capacity that may be used for the detector.  The way that the gain mechanism works inside an EMCCD adds noise to the signal which effectively reduces the detective quantum efficiency of the detector by a factor of~2.  The essentially zero readout noise of an EMCCD does allow it to work when the background sky brightness is very low, particularly at higher frame rates and when working at shorter wavelengths than $I$-band.  A major disadvantage o a detector design with EMCCDs is that approximately only 1/6 of the package area is taken up by sensitive silicon.  That means that the filling efficiency when using those detectors is relatively poor. In principle image slicers might be
used but these will be complicated, expensive and could only go some way
towards improving this fill factor.

\begin{figure}
\centering
\resizebox{\columnwidth}{!}{\includegraphics{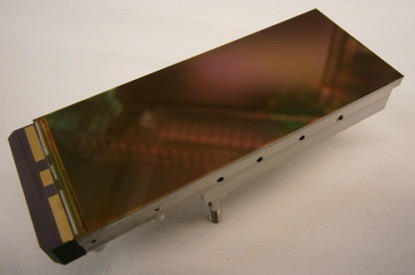}}
 \caption{Example of large area CMOS device, the CIS113 from Teledyne E2V (Chelmsford, UK) \citep{Jorden+2014}.  This has $1920 \times 4608$ pixels, each $16 \times 16~\umu\mbox{m}$.  It is back illuminated and 3-edge buttable.  This device has analogue outputs but other designs are available 
 with integrated signal processing electronics delivering very low read-out noise.  This device is approximately $30 \times 80~\mbox{mm}$.  Larger area devices may be made and they may also be constructed with the signal processing channels integrated onto the CMOS detector. This makes the driving and setup of the detector much easier. Multiple output channels are essential in order to achieve the frame rates required on large area detectors (see Figure~\protect\ref{fig:CMOS_scheme}).}
\label{fig:CMOS}
\end{figure}

\begin{figure}
\centering
\resizebox{4cm}{!}{\includegraphics{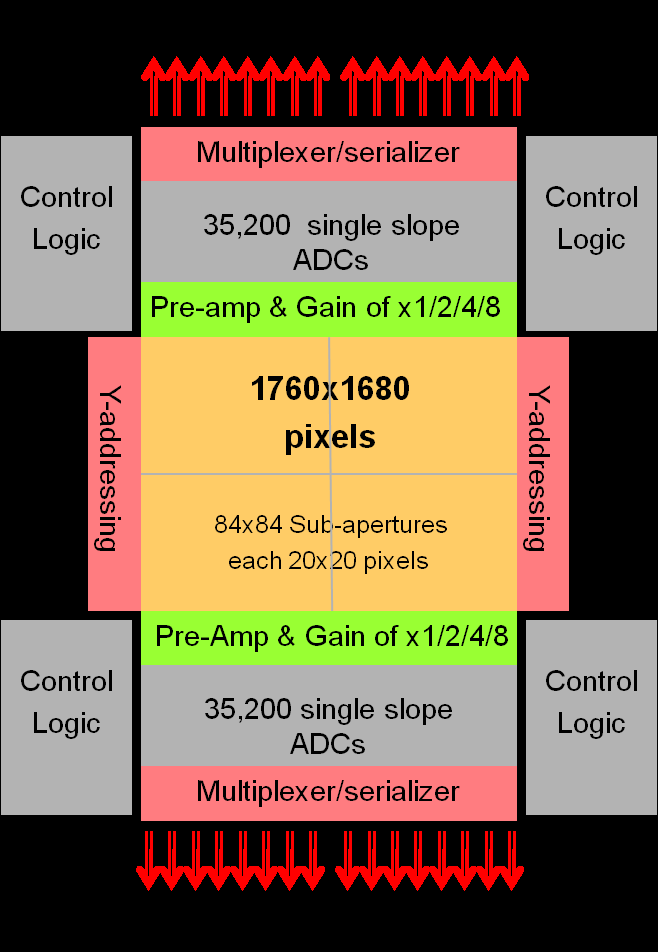}}
 \caption{Example of the internal organisation of a highly integrated CMOS sensor for astronomy.  The internal architecture is sub- divided into separate areas that may be read out in parallel.  This particular device is divided into sub apertures of $20 \times 20$~pixels, which may be read out randomly or sequentially.  The signals from the sub apertures are pre-amplified and then passed in this device to 70,400 single slope analogue to digital converters.  The digitised data are multiplexed to a total of 88~low-voltage differential serial links to be passed back to the computer \citep{Downing+2014}.}
\label{fig:CMOS_scheme}
\end{figure}

CCDs that do not have an amplification process will generally produce very high readout noise levels when operated at the frame rates we need.  Recently, considerable progress has been made with the development of high quantum efficiency CMOS detectors with very low read-out noise, $<1$~electron rms \citep{Segovia2017}. In principle they have a number of advantages.  The EMCCD uses a frame transfer architecture which means that the signal integrated in the sensitive part of the silicon is transferred rapidly at the end of each exposure into the storage area, where it is read out sequentially.  During the transfer of charge the device is still sensitive and bright objects in the field will produce a faint trail of the image that can complicate the photometry. Electronic
shutters might work in principle but with the fast frame readout the
device is working continually so the overall efficiency of the system
would be severely compromised. CMOS devices do not suffer from this (Figure~\ref{fig:CMOS}).  They use active pixel architecture which stores the charge and then, on command, transfers that signal into electronic components buried underneath the sensitive silicon.  While that charge is being read out the active pixel will integrate light for the following frame.  Astronomical CCDs have low read-out noise because they are read out very slowly.  CMOS detectors may be made with integrated signal processing electronics within the device itself.  Using one analogue processing chain with a single analogue to digital converter for each column of the detector, for example, means that each pixel is read out relatively slowly and excellent readout noise may be obtained (Figure~\ref{fig:CMOS_scheme}).  Readout noise levels below one electron rms have been obtained routinely in a number of devices.  This dramatically simplifies the driver electronics and greatly reduces power dissipation in the
vacuum enclosure.  CMOS technology is what is used within computer processor chips and therefore integrating even rather complicated electronics in a small area is relatively straightforward.  The latest CMOS devices are available with back illuminated (thinned) architectures and can be made with deeply depleted silicon that allows much higher contribution to the response in the far red part of the spectrum.  All these capabilities add significantly to the sensitivity and efficiency of the detector package.

The data from each of the columns of such a device is multiplexed in order to be transmitted back to the processing computer system.  Although this all sounds very complicated these processes may be substantially integrated and carried out within the CMOS device leading to a very simple package as far as the engineer employing such devices is concerned.  Such a device is driven digitally rather than with the more demanding analogue ones making the implementation of the \mbox{GravityCam} detector package significantly more straightforward.  Finally, it is also important that CMOS devices are now being manufactured so that they may be butted on 2 or~3 out of the 4~sides. Consequently, 85\% of the field of view of the telescope could be covered with a single pointing. This leads to considerable improvement in efficiency compared to the 16\% possible with a mosaic of EMCCDs, requiring 6 separate pointings to cover the full field of view. The intrinsic structure of the EMCCD makes it very hard to achieve fast buttable detectors without a very expensive custom development programme.  However it must be recognised that EMCCDs suitable for \mbox{GravityCam} already are available commercially. An appropriate CMOS detector would have to be developed but it is clear that the current technology levels would allow that to be done with some confidence. In particular, CMOS detectors still need to be characterised for reliable pixel-to-pixel photometry necessary for astronomical work, but studies on this are ongoing. An excellent summary of the current performance levels achieved by CMOS detectors is provided by \citet{Janesick2017}. 

Although the read noise per pixel from a CMOS detector is on average
around one electron RMS, some pixels exhibit significantly higher noise.
The nature of the turbulence being studied, however, means that the
centroid of the images most by approximately the half width of the
seeing profile, typically $0.7\arcsec$ equivalent to 9 pixels with
\mbox{GravityCam}, an area of 63 pixels. This means that the contribution to
the noise from any noisy pixel is greatly attenuated by the random
motion of the seeing disc once very many images are accumulated. Our
calculations indicate the net effect will be essentially negligible.

The key performance requirement is low readout noise at the same time as a high frame rate.  It is worth noting in passing that there are now also near infrared detectors manufactured by Selex \citep{Finger+2016} that are getting very close to the desired performance.  These use mercury cadmium telluride (MCT) sensors and cover the spectral range from about $0.8~\umu\mbox{m}$--$2.2~\umu\mbox{m}$
\citep{Hall+2016}.

\section{Computer interface and software structure}

Working on an assumption of $16~\umu\mbox{m}$ pixels and a nearly full focal plane, we will have to process about 10~Gpixels/sec assuming that the detectors operate at 25~Hz.  Experience with fast imaging detectors at Cambridge suggests that it will be relatively straightforward to manage at least 200 Mpixels/sec at 30~Hz frame rate with a single data processing computer.  We have simply assumed that each computer will be a relatively standard well-configured PC type computer.  Matching the data processing requirements, the entire detector array would be serviced by around 50 data processing PCs plus a small number of supervisory units charged with synchronising and overseeing the instrument. Digitised data from each module will be sent to the computer via optical fibres. The total data volume produced of $\sim\,400$~TB/night (one pixel delivering 2 bytes) is too large to transmit from the mountain, and that a thorough and reliable real-time processing pipeline is therefore as an important part of the instrument as the hardware.
  The simplest approach is to take each image, determine the relative offset of that image, then shift it and add it to the summed image.  In practice the atmospheric seeing is very variable and therefore it makes more sense to evaluate the sharpness of each image and sum those images into bins corresponding to a range of seeing conditions.  Images may be analysed individually but the data volume that produces would quickly become unmanageable.

As the point spread function is very well defined, when undertaking a gravitational microlensing survey there may be good reason to deconvolve each of the images so that PSF haloes are suppressed.  This degree of processing requires more significant provision and our models suggest that this may be done with graphics processor unit (GPU) cards or with many core
processors. Indeed, if they are to be used it is likely that much of the routine processing is also best done on those cards as the data will be transferred into them anyway.  

Our view of the way \mbox{GravityCam} would be principally used is for surveys which rely on accumulating a knowledge of the photometric characteristics of all the objects in each field of view.  Whenever a particular field is revisited by \mbox{GravityCam} there will be in the archive a detailed existing knowledge of that field which will be loaded into the computer.  Each new image can be compared immediately with the knowledge base that exists already for that field.  In the case of the gravitational microlensing survey, the principal data needed for each star is an understanding of its intrinsic variability.  When looking for gravitational microlensing events, the first sign that an event is starting will be a star increasing in brightness in a manner that is very different from its usual behaviour.  Such an event must be flagged immediately to ensure that the field is revisited frequently enough to provide the photometric monitoring as necessary.  This enables other instruments or telescopes programs to follow the new event as appropriate.  Indeed it would also be possible to change the observing  program of \mbox{GravityCam} to provide follow-up photometry.
 
 The management of \mbox{GravityCam} has to be done by a supervisory system which is responsible for  the setup, characterisation and testing of each of the detectors in the camera, recording performance information as appropriate, managing data archiving and backup and other management functions.  Once all the detectors are running and operating at the designated temperature data taking may be started.  For each new field accessed, where appropriate, archival data on that field is loaded into each data processing computer.  The supervisory system triggers each data processing computer to proceed with the observations of each field.  Each image is calibrated photometrically simply by measuring the light from as many reference stars as available to give a photometric calibration of each individual frame.  Stars are then tracked and their brightness compared with the existing knowledge of their characteristics.  Individual images are then offset laterally to bring them in synchronism with the established images.  The new images may then be combined with images already summed , and processed to improve the detailed knowledge of the characteristics of each and every star in the field, in turn allowing the archive data to be improved and updated.  The exact frequency with which this is done depends very much on a detailed trade-off involving the processing power required and the information needed and of the particular observing program.
 
 \section{Sensitivity estimates and predictions}
 \label{Sect:sensitive}
 The basic concept of image selection does not reduce the efficiency or sensitivity of an imaging process in any way relative to the sensitivity that might be achieved with single long exposure.  This would not be true if the background signal level was very low and readout noise dominated after many frames had been integrated.  With 100\% image selection followed by shifting and adding then there is no loss in efficiency.  Neither of these detectors (EMCCD or CMOS) has essentially any equivalent of "shutter closed time".  In practice it may be that individual images are accumulated into more specific image quality bins such as 10\% windows.  These can always be combined later to give the full efficiency.  However if only 50\% of the images are to be used then the total signal in the final image will be reduced by a factor of~2.
 
 It has already been mentioned that gravitational lensing, both microlensing and weak shear lensing, is an achromatic process.  This means that in order to achieve the very highest sensitivity then filters may be dispensed with or perhaps restricted to a long pass filter.  It is very unfashionable for astronomers to observe without filters but it has a significant influence on the overall sensitivity.  With typical CCD or CMOS sensitivity, dispensing with a filter will improve the signal-to-noise on a faint object by about one magnitude when compared with the I band sensitivity.  It is interesting to note that deep galaxy imaging by \citet{HalMac1984} without any filter detected high galaxy surface densities at high galactic latitudes.  Higher densities were only measured nearly 20 years later by the Hubble Space Telescope Deep Field and by using very long exposures indeed.  Moreover, the Euclid space mission scheduled for launch in 2020 (http://www.euclid-ec.org/)  includes an imaging instrument (VIS) with only a very broad band (500--900~nm) filter for just this purpose.  As a small diameter telescope it is important that it has as high a throughput as possible.
 
 
 
With EMCCDs that are already developed we know that the cost of covering a 0.5$^\circ$ diameter focal plane will be in the region of \$1~million. With CMOS detectors the costs are less easily quantified because the devices we need have to be developed. However there is no doubt that devices with the right specification for GravityCam can be made by using existing CMOS technology already developed by Teledyne E2V. This development is relatively low risk. Also, if we use CMOS detectors, we will be essentially purchasing 6 times the sensitive area and the cost of silicon detectors that are thinned will be correspondingly higher. At this stage the cost of these detectors will clearly drive the total cost and so accurate estimates cannot really be made. An approach to start with EMCCDs and upgrade to CMOS devices at a later date would be possible although the aggregate cost would therefore be significantly greater. EMCCD and CMOS driver electronics are very different so they would also have to be reworked. We estimate that the project cost using CMOS detectors might be in the range of \$12--\$17~million. These costs include allowing one year for commissioning plus a further~2 years of continuing support. Most of the work required is relatively straightforward and could be done within 3~years. This makes the GravityCam project relatively inexpensive and relatively quick to implement on the telescope. It has the potential to revolutionise several independent branches of astronomy and provides a unique capability not available on any other telescope, ground or space based or indeed planned in the foreseeable future.

\section{Conclusions}
\label{Sect:Conclusions}

There is clearly a very strong scientific case to make a wide-field survey instrument that can deliver much sharper images on ground-based telescopes.  Such an instrument would revolutionise our understanding of many aspects of planet formation by permitting detections of planets and satellites down to Lunar mass across the Milky Way.  The quality of data that may be taken for weak shear gravitational lensing studies will be significantly better than is possible otherwise. Studies of fast multiwavelength flaring in accreting compact objects can be carried out. Moreover, the datasets created by \mbox{GravityCam} will have considerable impact on the development of asteroseismology and its capacity to detect Kuiper belt objects and possibly even objects from the Oort cloud is very exciting indeed.

Developments in imaging detector technology mean that techniques of improving the resolution of images taken on ground-based telescopes have been thoroughly demonstrated and are in use already on a number of ground-based telescopes.  All the principles of building a wide-field instrument have been demonstrated already and there are programs which are now being substantially constrained because of the quality of images that can be delivered even on the best ground-based sites.  \mbox{GravityCam} provides a new approach to how this can be done and we are confident the \mbox{GravityCam} has a very great potential as a new class of astronomical survey instrument.

\section*{Acknowledgements}

We have been very fortunate in having help and advice from a number of people.  In particular, Lindsay King, Sarah Bridle and Michael Hirsch have contributed to important sections of this paper.  We have also benefited from discussions with Tom Kitching, Stuart Bates, Rafael Robolo, Tim Staley, Keith Horne, Bernhard Sch\"{o}lkopf, Laurent Gizon, Alexandre Refregier, Adam Amara, Don Pollacco, Cesare Barbieri, and Giampiero Naletto.
  We are also grateful for help and advice by the staff of the observatory operated by the European Southern Observatory at La Silla, Chile. We would like to thank SPIE for granting permission to reprint material previously presented by \citet{SPIE:GravityCam}.


\bibliographystyle{mnras}
\bibliography{GravityCam_pasa.bib}

\end{document}